\RequirePackage{lineno}
\documentclass[aps,prl,twocolumn,floatfix,superscriptaddress,balancelastpage,footinbib,amsmath,preprintnumbers]{revtex4-1}[11pt]
\pdfoutput=1
\usepackage{graphicx}
\usepackage{amsmath,amssymb,mathrsfs}
\usepackage{bbm}
\usepackage{color}
\usepackage{xcolor}
\usepackage{dsfont}
\usepackage{cancel}
\usepackage{dsfont}
\usepackage{epstopdf}
\usepackage{epsfig}
\usepackage{bm}
\usepackage{dcolumn}
\usepackage{hyperref}

\usepackage[utf8]{inputenc}

\newcommand{\ben}{\begin{enumerate}}
\newcommand{\een}{\end{enumerate}}
\newcommand{\bit}{\begin{itemize}}
\newcommand{\eit}{\end{itemize}}

\newcommand{\beqa}{\begin{eqnarray}}
\newcommand{\eeqa}{\end{eqnarray}}
\newcommand{\beq}{\begin{equation}}
\newcommand{\eeq}{\end{equation}}
\newcommand{\bay}{\begin{array}}
\newcommand{\eay}{\end{array}}

\def\ifmath#1{\relax\ifmmode #1\else $#1$\fi}

\def\gsim{\ \rlap{\raise 3pt \hbox{$>$}}{\lower 3pt \hbox{$\sim$}}\ }
\def\lsim{\ \rlap{\raise 3pt \hbox{$<$}}{\lower 3pt \hbox{$\sim$}}\ }

\def\ls#1{\ifmath{_{\lower1.5pt\hbox{$\scriptstyle #1$}}}}
\def\lsup#1{^{\lower 6pt\hbox{$\scriptstyle#1$}}}

\def\bracket#1#2 {\mathinner{\langle{#1}|{#2}\rangle}}

\def\bracket#1#2 {\mathinner{\langle{#1}|{#2}\rangle}}

\newcommand{\mysection}[1]{{\noindent \bf {#1.}}}

\newcommand{\bea}{\begin{eqnarray}}
\newcommand{\eea}{\end{eqnarray}}

\graphicspath{{figs/}}
\begin{document}


\title{New Constraints and Prospects for sub-GeV Dark Matter \\ Scattering off Electrons in Xenon}

\author{Rouven Essig}
\email{rouven.essig@stonybrook.edu}
\affiliation{C.N. Yang Institute for Theoretical Physics, Stony Brook University, Stony Brook, NY 11794}

\author{Tomer Volansky}
\email{tomerv@post.tau.ac.il}
\affiliation{Raymond and Beverly Sackler School of Physics and Astronomy, Tel-Aviv University, Tel-Aviv 69978, Israel}

\author{Tien-Tien Yu}
\email{tien-tien.yu@cern.ch}
\affiliation{C.N. Yang Institute for Theoretical Physics, Stony Brook University, Stony Brook, NY 11794}
\affiliation{Theoretical Physics Department, CERN, CH-1211 Geneva 23, Switzerland}

\preprint{YITP-SB-17-09, CERN-TH-2017-042}

\begin{abstract}
We study in detail sub-GeV dark matter scattering off electrons in xenon, including the expected electron recoil spectra and annual 
modulation spectra.  
We derive improved constraints using low-energy XENON10 and XENON100 ionization-only data. 
For XENON10, in addition to including electron-recoil data corresponding to about $1-3$ electrons, we include for the first time events 
with $\gtrsim 4$ electrons.  
Assuming the scattering is momentum independent, this strengthens a previous cross-section bound by almost an order of magnitude 
for dark matter masses above 50~MeV.  
The available XENON100 data corresponds to events with $\gtrsim 4$ electrons, and leads to a constraint that is comparable to 
the XENON10 bound above 50 MeV. 
We demonstrate that a search for an annual modulation signal in upcoming xenon experiments (XENON1T, XENONnT, LZ) 
could substantially improve the above bounds even in the presence of large backgrounds.  
We also emphasize that in simple benchmark models of sub-GeV dark matter, 
the dark matter-electron scattering rate can be as high as one event every ten (two) seconds in the 
XENON1T (XENONnT or LZ) experiments, 
without being in conflict with any other known experimental bounds.  
While there are several sources of backgrounds that can produce single- or few-electron events, 
a large event rate can be consistent with a dark matter signal and should not be simply written off as purely a detector curiosity.  
This fact motivates a detailed analysis of the ionization-only (``S2-only'') data, taking into account 
the expected annual modulation spectrum of the signal rate,
as well as the DM-induced electron-recoil spectra, 
which are another powerful discriminant between signal and background.    
\end{abstract}

\maketitle

\mysection{Introduction}  
Direct-detection experiments play a crucial role in our quest to identify the nature of dark matter (DM), and 
the last few years have seen intense interest and significant progress in expanding their sensitivity 
to particles below $\sim 1$~GeV.  
The traditional direct detection technique --- observing nuclear recoils from DM scattering elastically 
off nuclei --- rapidly loses sensitivity in existing experiments for DM masses below $\sim 1$~GeV, calling for different approaches.  
A demonstrated technique with significant potential for improvement is to search for DM scattering off electrons~\cite{Essig:2011nj}. 
Various target materials have been investigated, including noble liquids~\cite{Essig:2011nj,Essig:2012yx}, semiconductors~\cite{Essig:2011nj,Graham:2012su,Lee:2015qva,Essig:2015cda}, scintillators~\cite{Derenzo:2016fse}, two-dimensional targets~\cite{Hochberg:2016ntt}, and superconductors~\cite{Hochberg:2015fth,Hochberg:2015pha}.  
These materials are also sensitive to the absorption of ultralight DM ($\ll$~MeV) by 
electrons~\cite{An:2013yua,An:2014twa,Bloch:2016sjj,Hochberg:2016sqx}.  
For other direct-detection ideas see~\cite{Essig:2011nj,Essig:2016crl,CCupcoming,Schutz:2016tid,Knapen:2016cue,Kouvaris:2016afs,McCabe:2017rln,Bunting:2017net}.  Direct-detection techniques and complementary probes are summarized in~\cite{Essig:2013lka}. 

Currently, the most stringent direct-detection constraint on DM as low as a few MeV comes from XENON10, a two-phase xenon time 
projection chamber (TPC).  
When a DM particle scatters off an electron and ionizes a xenon atom in the liquid target, 
the recoiling electron can ionize other atoms if it has sufficient energy.  
An electric field accelerates the ionized electrons through the liquid, across a liquid-gas interface, and through a xenon gas region in which interactions between the electrons and xenon atoms create a scintillation (``S2'') signal that is 
proportional to the number of extracted electrons and detected by photomultiplier tubes.  
XENON10~\cite{Angle:2011th} has taken data consisting of events that have an S2 signal corresponding to one or more electrons, 
without an observable prompt scintillation signal (``S1'').  
The data corresponding to events with three electrons or less ($n_e\lesssim 3$) were analyzed in~\cite{Essig:2012yx} and 
shown to constrain DM as low as a few MeV.  

The main factor limiting the sensitivity of XENON10 is the large number of observed S2-only events and the absence of 
a background model (to set a constraint, all events are conservatively assumed to originate from DM).  
Plausible origins of these events include 
the photo-dissociation of negatively charged impurities; ionized electrons that were initially created by highly ionizing background events, but then became trapped in the liquid-gas interface and spontaneously released at a later time; and field emission from the 
cathode~\cite{Angle:2011th,Essig:2012yx}.  
However, more study is needed to understand and characterize these events.  

In this letter, we derive new constraints from XENON10, including events with $n_e\gtrsim 4$.  The rate of such events is lower than 
for $n_e\lesssim 3$, leading to significantly improved constraints for DM masses $m_\chi \gtrsim 50$~MeV. 
We also analyze S2-only data from XENON100, containing $n_e\gtrsim 4$~\cite{Aprile:2016wwo}. 
We derive the expected recoil spectra for the event rate and the annual and daily modulation amplitude, and show 
the expected event rates and the implications for a few benchmark DM models.  

Other experiments, using semiconductor targets such as germanium (Ge) and silicon (Si), currently have a higher electron-recoil energy threshold and are thus less sensitive by several orders of magnitude than XENON10/100~\cite{Essig:2015cda}.  
Dramatic improvements in sensitivity in the near future are likely with SuperCDMS~\cite{Cushman:2013zza, Essig:2015cda}, SENSEI~\cite{SENSEI}, and possibly other experiments. 
Nevertheless, these experiments will initially have target masses of only $\mathcal{O}$(1~kg), far less than current and 
future xenon experiments (Table~\ref{tab:exposures}).  Understanding the S2-only events in two-phase TPCs could thus  
lead to dramatic improvements in cross-section sensitivity and, as we will show, probe simple and predictive benchmark models.  
The large exposures will also allow for an annual modulation analysis~\cite{Drukier:1986tm}, 
which can significantly improve upon the current limit even if the background rates are high.  

\begin{table}[t]
\begin{center}
\begin{tabular}{|c||c|c|}
\hline
 & exposure [kg-yrs] & fiducial mass [kg] \\ \hline
XENON10~\cite{Angle:2011th} & 0.041 & 1.2\\ \hline
XENON100~\cite{Aprile:2016wwo} & 29.8 & 48.3\\ \hline
LUX~\cite{Akerib:2016lao} &  119 
					& 145\\ \hline 
XENON1T~\cite{Aprile:2015uzo} & 2,000 & 1,000 \\ \hline 
LZ, XENONnT~\cite{Akerib:2015cja,Aprile:2015uzo} & 15,000  
							& 5,600\\ \hline
\end{tabular}
\caption{Analyzed (XENON10, XENON100, LUX) and approximate projected (XENON1T, LZ, XENONnT) exposures and fiducial masses. \vspace{-6mm}
}
\label{tab:exposures}
\end{center}
\end{table}%

\vspace{1mm}
\mysection{Theoretical Rates and Recoil Spectra} 
To calculate the DM-electron scattering rate in liquid xenon, we follow the procedure in~\cite{Essig:2012yx} 
(see appendices for more details).  
We treat the target electrons as single-particle states of an isolated atom, 
described by numerical RHF bound wave functions from~\cite{BUNGE1993113} \footnote{Experimental results may suggest that liquid xenon has a band structure with a gap of about 9.2~eV~\cite{PhysRevB.10.4464}.  In this case, our calculation, which assumes an ionization energy of 12.4~eV, may underestimate slightly the true scattering rate.}. 
The velocity-averaged differential ionization cross section for electrons in the $(n,l)$ shell is 
\beq\label{eq:ionization-sigma}
\frac{d\langle\sigma_{ion}^{nl}\rangle}{d\ln E_{er}}=\frac{\overline\sigma_e}{8\mu_{\chi e}^2}\int qdq |f_{ion}^{nl}(k',q)|^2|F_{\rm{DM}}(q)|^2\eta(v_{min}),
\eeq
where
$\eta(v_{min})=\langle\frac{1}{v}\theta(v-v_{min})\rangle$ is the inverse mean speed for a given velocity distribution as a function of the minimum velocity, $v_{min}$, required for scattering.   We assume a standard Maxwell-Boltzmann velocity distribution with circular velocity $v_0=220$ km/s and a hard cutoff of $v_{esc}=544$ km/s ~\cite{Smith:2006ym,Dehnen:1997cq}. $\overline\sigma_e$ is the DM-free electron scattering cross section at fixed momentum transfer $q=\alpha m_e$, while the $q$-dependence of the matrix element is encoded in the DM form-factor $F_{\rm{DM}}(q)$.  $|f_{ion}^{nl}(k',q)|^2$ is the ionization form factor of an electron in the $(n,l)$ shell with final momentum $k'=\sqrt{2m_e E_{er}}$.  We calculate this form factor using the given bound wave functions and unbound wave functions that are obtained by solving the Schr\"odinger equation with a potential that reproduces the bound wave functions.  We consider electrons in the following shells (listed with binding energies in eV): 
$5p^6$ (12.4), $5s^2$ (25.7), $4d^{10}$ (75.6), $4p^6$ (163.5), and $4s^2$~(213.8).  
The differential ionization rate is  
\beq
\frac{dR_{ion}}{d\ln E_{er}}=N_T\frac{\rho_{\chi}}{m_{\chi}}\sum_{nl}\frac{d\langle \sigma_{ion}^{nl}v\rangle}{d\ln E_{er}},
\eeq
where $N_T$ is the number of target atoms and $\rho_\chi=0.4$~GeV/cm$^{3}$ is the local DM density.

\begin{figure}[t]
\vskip -4mm
\includegraphics[width=0.46\textwidth]{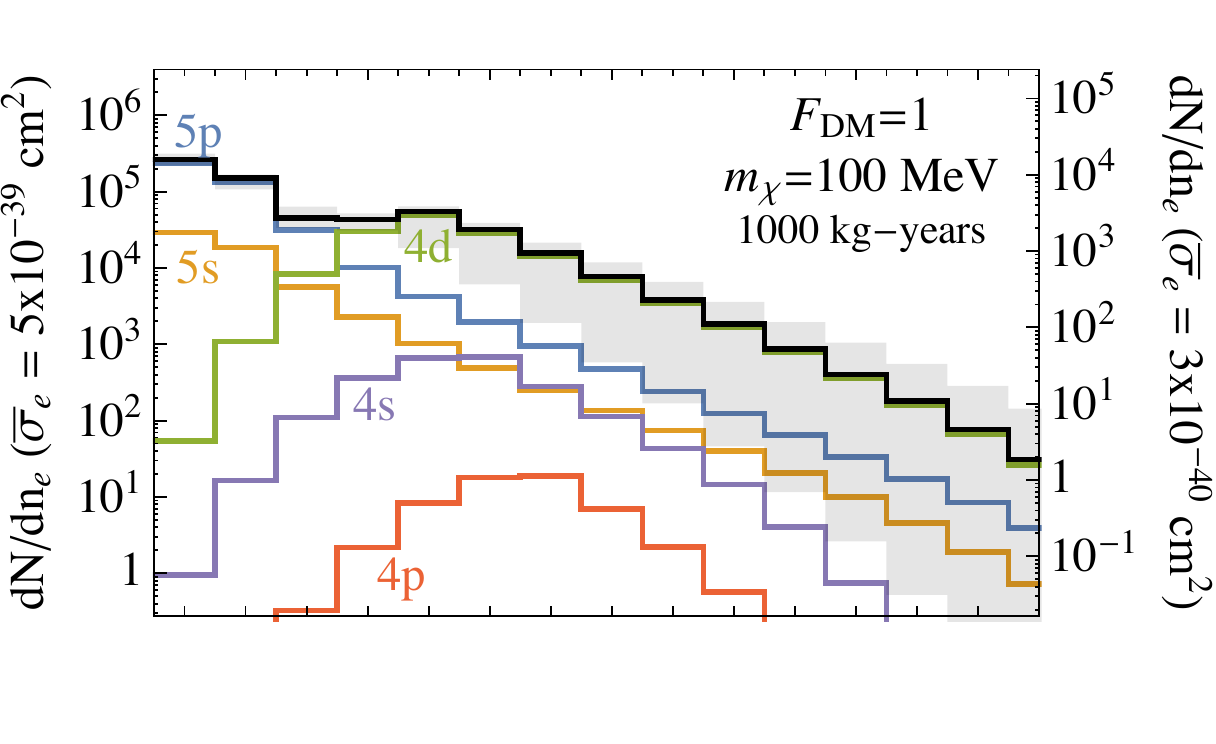}\\
\vskip -13.7mm
\hskip -0.5mm 
\includegraphics[width=0.472\textwidth]{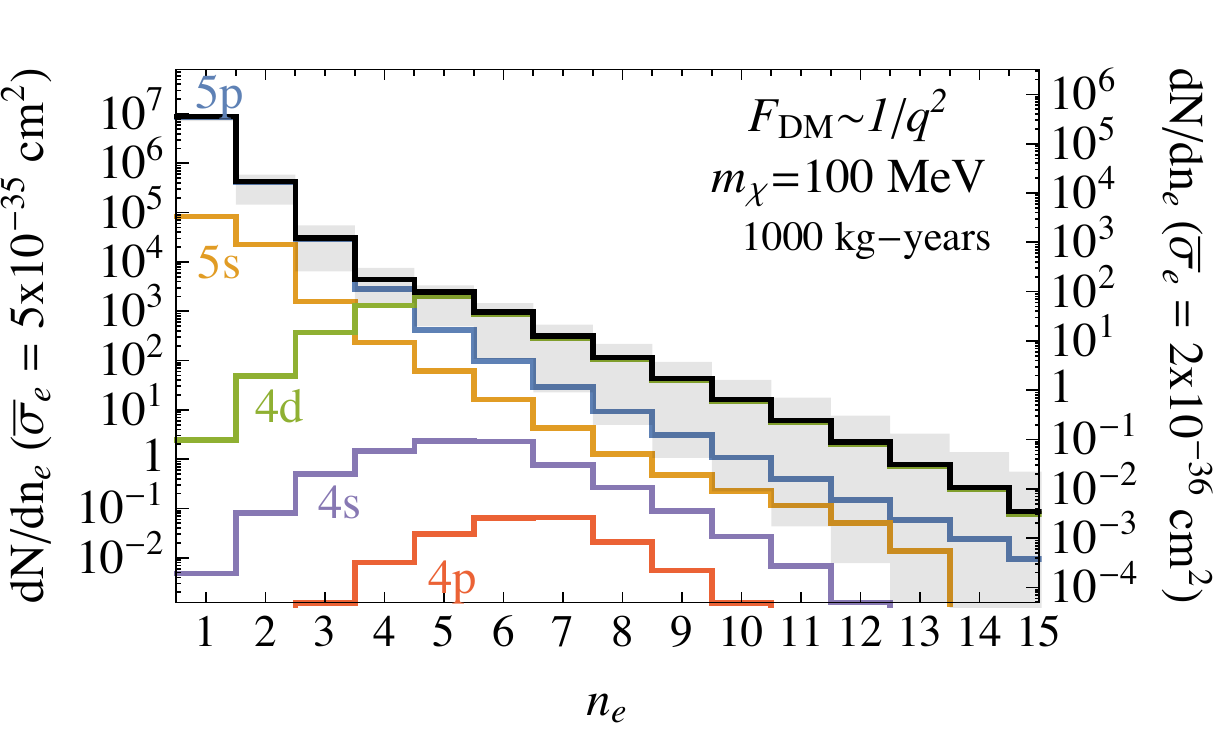}
\vspace{-3mm}
\caption{
{\bf Top} ({\bf bottom}): 
Spectrum of expected number of events for DM-electron scattering in xenon, for $m_\chi=100$~MeV and 1000 kg-years for 
$F_{\rm DM}=1~(\alpha^2 m_e^2/q^2)$. 
For the left axes, we set $\overline\sigma_e$ to the maximum allowed values by current constraints for two popular benchmark models; 
for the right axes, the indicated $\overline\sigma_e$ produces the correct relic abundance. 
Colored lines show individual contributions from various xenon electron shells, while the gray band encompasses the spectrum 
when varying the secondary ionization model.  See text for details. 
\vspace{-3mm}
}
\label{fig:spectramain}
\end{figure}
 
\begin{figure}[t]
\hskip -2mm
\includegraphics[width=0.467\textwidth]{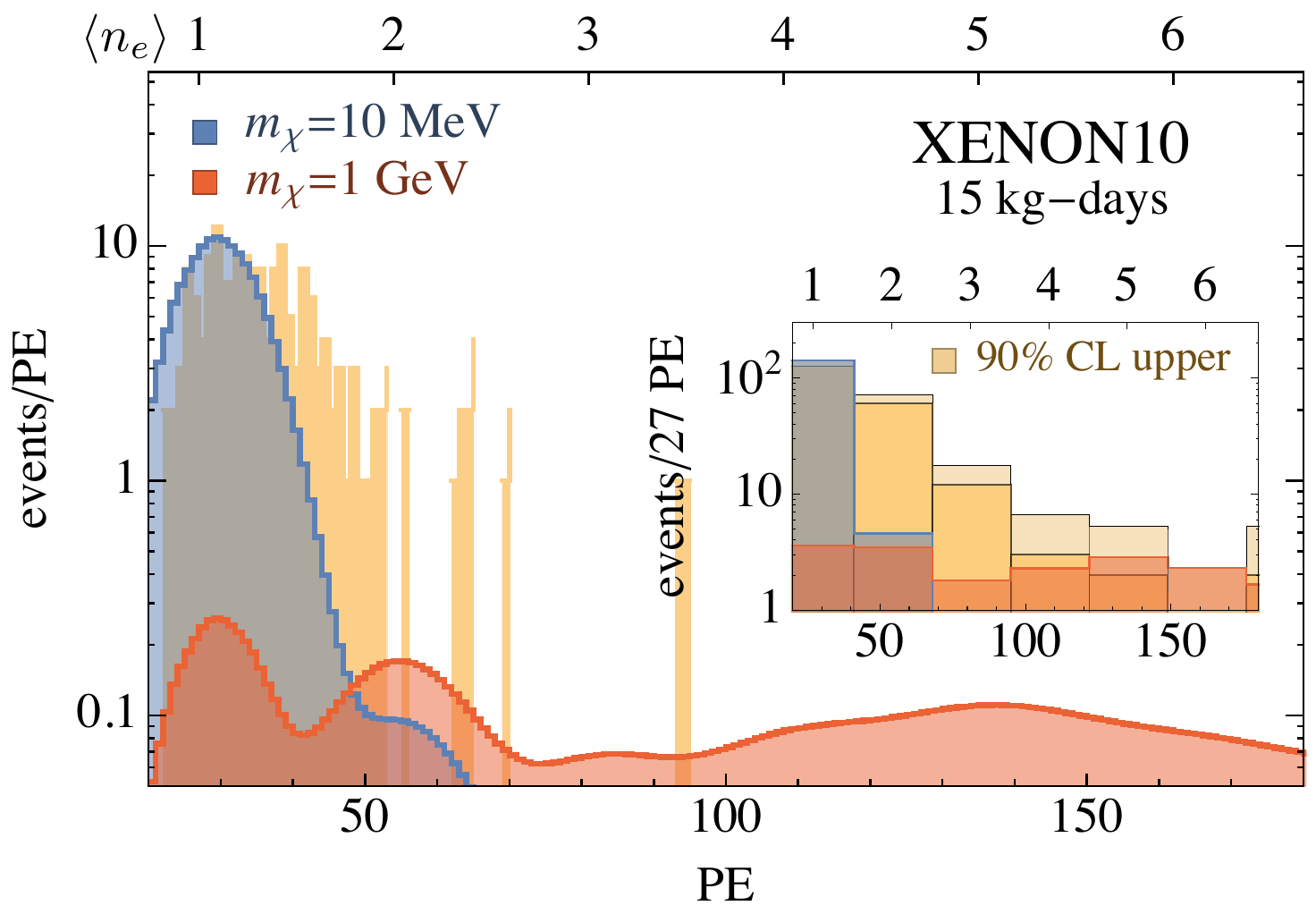}\\
\includegraphics[width=0.47\textwidth]{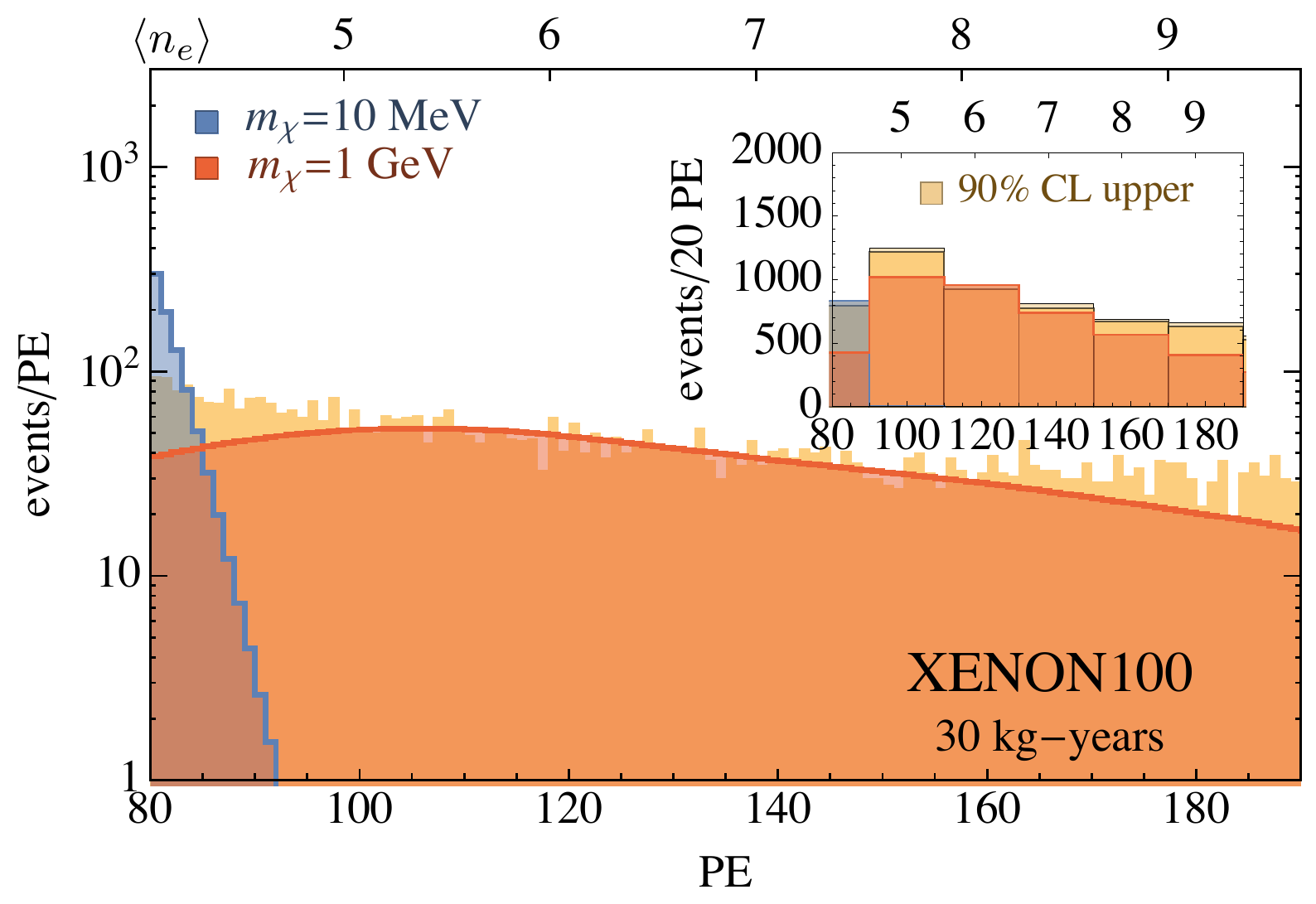}
\caption{Observed number of events versus photoelectrons (PE) in XENON10 ({\bf top})~\cite{Angle:2011th} 
and XENON100 ({\bf bottom})~\cite{Aprile:2016wwo}.  
DM spectra are shown for $m_\chi=10$~MeV (blue) \& 1~GeV (red)  with a cross section fixed at our derived  
90\% C.L.~limit (we assume fiducial values for the secondary ionization model). 
Insets show spectra in bins of 27PE (20PE), the mean number of PE created by one electron in XENON10 (XENON100).
}
\label{fig:sample-spectra}
\end{figure}

We follow~\cite{Essig:2012yx} to model the conversion from $E_{\rm er}$ to electron yield, $n_e$.   
The recoiling electron will ionize and excite other atoms, producing $n^{(1)} = {\rm Floor}(E_{\rm er}/W)$ {\it additional} ``primary quanta'', 
either observable electrons or (unobservable) scintillation photons.  
For fiducial values, we choose the probability for the initial electron to recombine with an ion to be $f_R=0$, $W=13.8$~eV, and 
the fraction of primary quanta observed as electrons to be $f_e = 0.83$.  
To capture the uncertainty in the fiducial values, we vary these parameters in the range $0 < f_R < 0.2$, 
$12.4 < W < 16$~eV, and $0.62 < f_e < 0.91$.  
In addition to primary quanta, if DM ionizes an inner-shell electron, $n^{(2)}={\rm Floor}((E_i-E_j)/W)$ secondary quanta can be created by photons produced in the 
subsequent outer-to-inner-shell electron transitions with binding energies $E_{i,j}$.  
The number of secondary electrons produced follows a binomial distribution with $n^{(1)}+n^{(2)}$ trials and success probability $f_e$.

In Fig.~\ref{fig:spectramain}, we show the recoil spectra as a function of $n_e$ for a hypothetical xenon detector with 1000 kg-years 
of exposure for $F_{\rm DM}=1$ (top) and $F_{\rm DM} = \alpha^2 m_e^2/q^2$ (bottom).  
The colored lines show individual contributions from different shells, while the black line shows their sum (for fiducial values).  
Gray bands show the variation away from the fiducial values discussed above. 

To emphasize the importance of studying electron recoil events at current and upcoming xenon experiments, we have fixed 
$\overline\sigma_e$ to specific values that are allowed by simple and predictive benchmark models~\cite{Boehm:2003hm,Borodatchenkova:2005ct,Batell:2009di,Essig:2011nj,Chu:2011be,Lin:2011gj,Izaguirre:2015yja,Essig:2015cda,Alexander:2016aln} 
and further below.  We consider the DM (a Dirac fermion 
or complex scalar $\chi$) to be charged under a broken $U(1)_D$ gauge force, 
mediated by a kinetically-mixed dark photon, $A'$, with mass $m_{A'}$.  
The $A'$ mediates DM-electron scattering, and $F_{\rm DM}(q) = 1$ $(\alpha^2 m_e^2/q^2)$ for a heavy (ultralight) dark photon.   
The left axis for top (bottom) plot of Fig.~\ref{fig:spectramain} shows the event rate for $\overline \sigma_e$ fixed to the maximum 
value allowed by current constraints for $m_{A'}=3m_\chi$ ($m_{A'}\ll~{\rm keV}$), 
while the right axis of the top (bottom) plot fixes $\overline \sigma_e$ so that scalar (fermion) DM obtains the correct relic abundance from  
thermal freeze-out (freeze-in).  
Clearly, a large number of DM events could be seen in upcoming detectors.     
These results are easily rescaled to other DM models that predict DM-electron scattering.

\begin{figure}[t]
\includegraphics[width=0.48\textwidth]{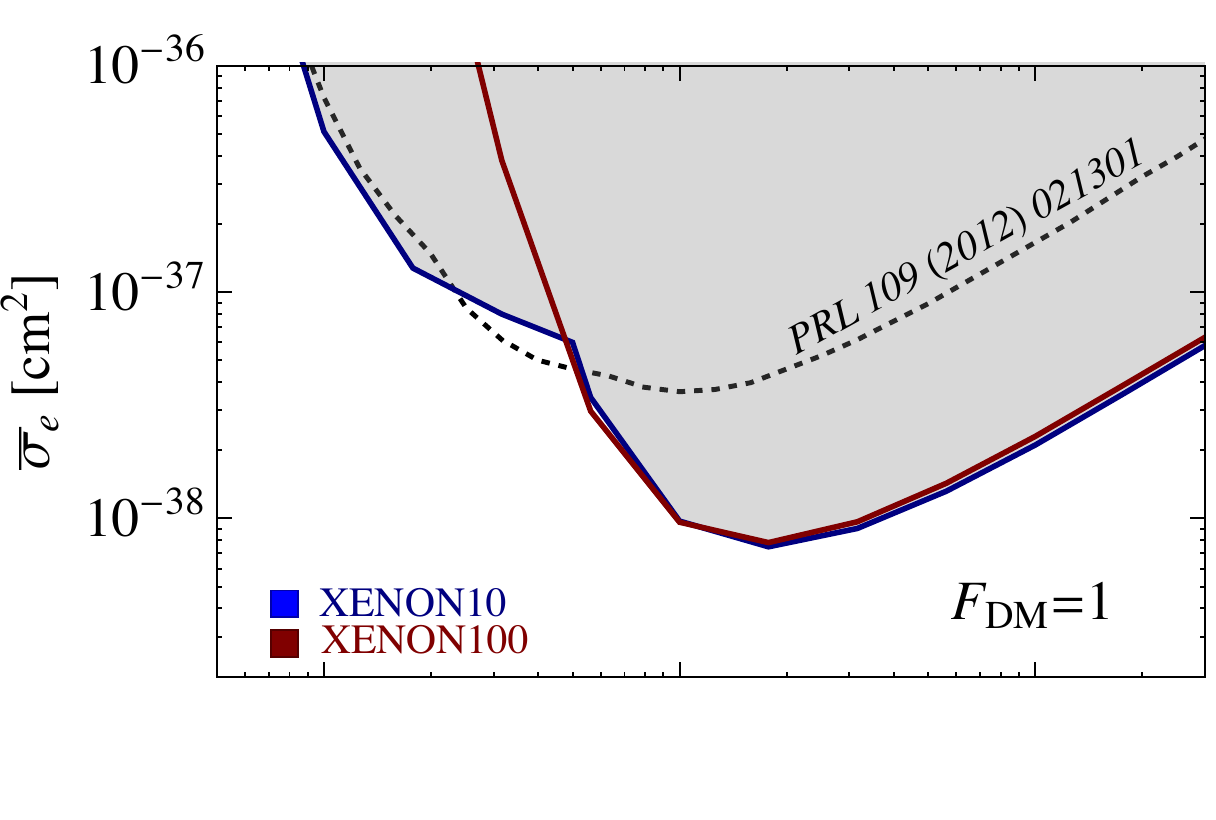}\\
\vskip -15mm
\includegraphics[width=0.48\textwidth]{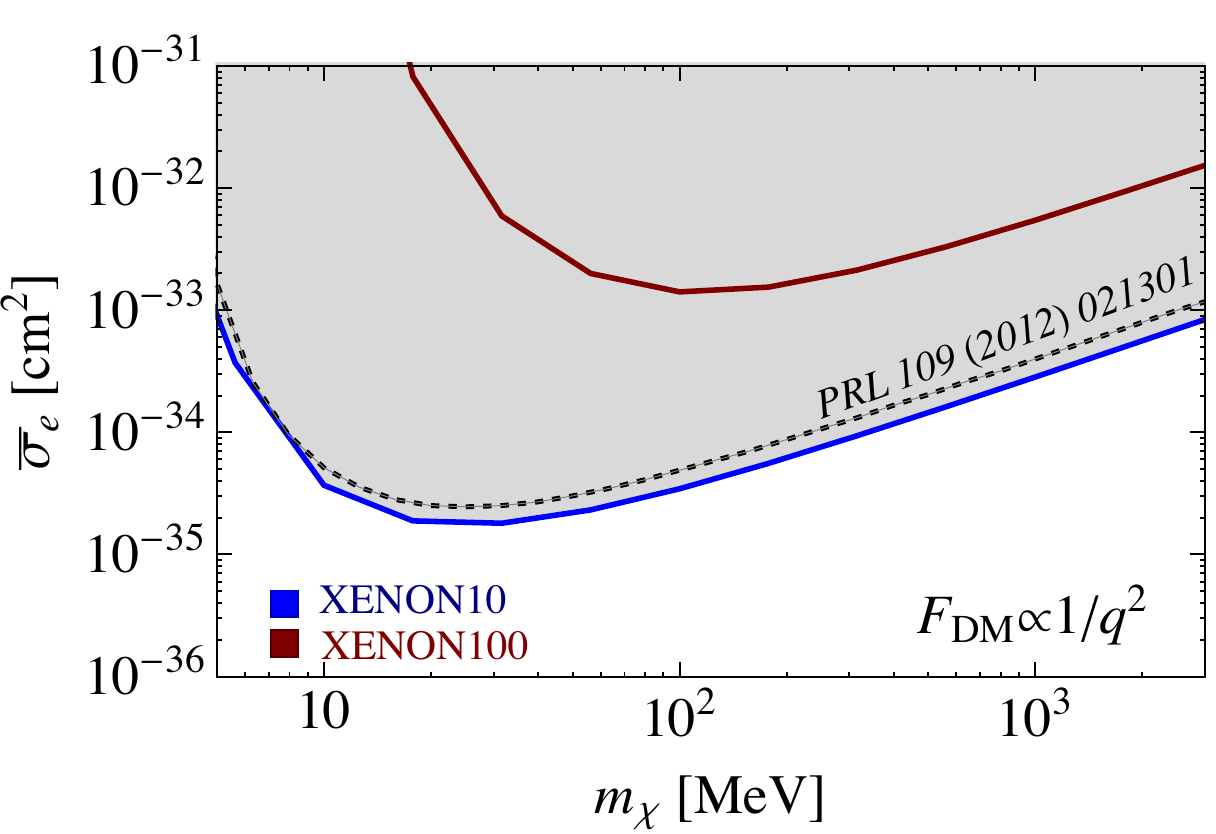}
\caption{90\%~C.L.~limit on the DM-electron scattering cross section from XENON10 data (blue) and XENON100 data (red) for 
$F_{\rm DM}=1$ ({\bf top}) \& $F_{\rm DM} = \alpha^2 m_e^2/q^2$ ({\bf bottom}).  
Dotted black lines show XENON10 bounds from~\cite{Essig:2012yx}.}
\label{fig:xenon10_vs_xenon100-combined}
\end{figure}

\mysection{New XENON10 and XENON100 bounds} 
We now recalculate the bounds from XENON10 data~\cite{Essig:2012yx} (15 kg-days), including for the first time events 
with $n_e\gtrsim 4$, as well as from XENON100 data~\cite{Aprile:2016wwo} (30 kg-years).  
Since the experimental observable is the number of photoelectrons (PE) produced by an event, we convert $n_e$ to PE.  
An event with $n_e$ electrons produces a gaussian distributed number of PE with mean $n_e \mu$ and 
width $\sqrt{n_e} \sigma$, where $\mu=27~(19.7)$ and $\sigma=6.7~(6.2)$ for XENON10 (XENON100). 
We multiply the signal with the trigger and acceptance efficiencies from \cite{Essig:2012yx,Aprile:2016wwo} and then 
bin both the signal and data in steps of 27PE (20PE), starting from 14PE (80PE) for XENON10 (XENON100). The first bin for the XENON100 analysis is 80-90PE, corresponding to roughly half an electron. We require that the resulting signal is less than the data at 90\%~C.L. in each bin.  
For XENON10, the 90\%~C.L.~upper bounds on the rates (after unfolding the efficiencies) are $r_1<15.18,~r_2< 3.37,~r_3<0.95,~r_4< 0.35,~r_5<0.35,~r_6<0.15,~r_7<0.35$  counts kg$^{-1}$ day$^{-1}$,
corresponding to bins $b_1=[14,41],~b_2=[41,68] \ldots, b_7=[176-203]$~PE; for XENON100, we find 
$r_4<0.17,~r_5<0.24,~r_6<0.17$ counts kg$^{-1}$ day$^{-1}$ corresponding to bins $b_4=[80,90], b_5=[90,110],~b_6=[110,130]$~PE. 

Fig.~\ref{fig:sample-spectra} shows the two data sets in PE and two sample DM spectra.  
Fig.~\ref{fig:xenon10_vs_xenon100-combined} shows the strongest XENON10 and XENON100 limit combined across all bins, and a  comparison with the XENON10 bound derived in~\cite{Essig:2012yx}.  In the Appendices, we show cross-section bounds for the individual PE bins, taking into account the systematic uncertainties from the secondary ionization model.
For $F_{\rm DM}=1$, the inclusion of the high-PE bins in XENON10 significantly improves upon the bound from~\cite{Essig:2012yx} for 
$m_\chi\gtrsim 50$~MeV (small differences at lower masses are from the limit-setting procedure). 
The new XENON10 and XENON100 bounds are comparable for $m_\chi\gtrsim 50$ MeV.  
For $F_{\rm DM} = \alpha^2m_e^2/q^2$, the low PE bins determine the bound, and XENON100 is therefore not competitive due to its high analysis threshold.

\mysection{Modulation} 
\begin{figure}[t]
\begin{center}
\includegraphics[width=0.46\textwidth]{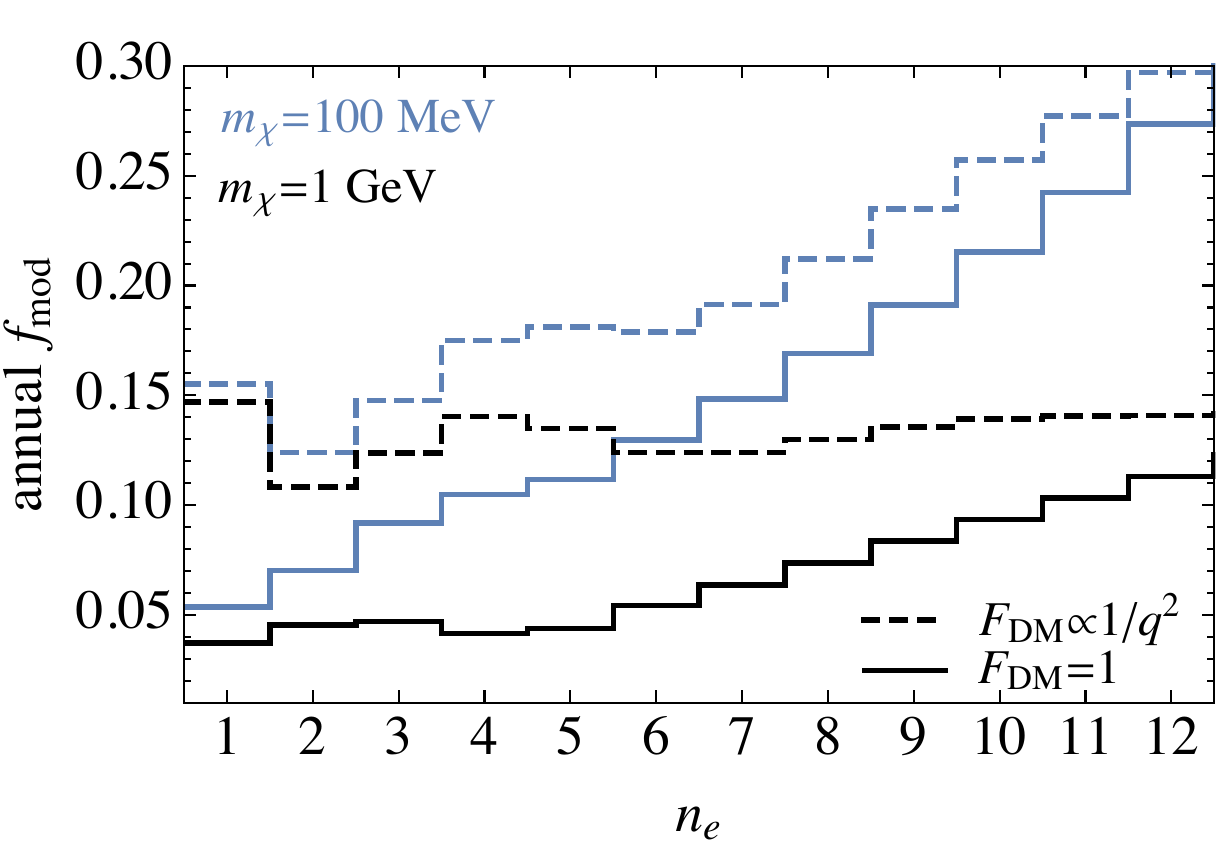}
\caption{Annual modulation amplitude for $F_{\rm DM}=1$ (solid) \& $F_{\rm DM} = \alpha^2 m_e^2/q^2$ (dashed) for 
$m_\chi=100$~MeV (blue) \& 1~GeV (black).
\vspace{-5mm}}
\label{fig:fmod_vs_Q}
\end{center}
\end{figure}
A useful discriminant between signal and background is the annual modulation of the signal rate~\cite{Drukier:1986tm} due to 
the Sun's motion through the DM halo. 
Fig.~\ref{fig:fmod_vs_Q} shows $f_{\rm mod}$ versus $n_e$, where 
$f_{\rm{mod}}=\frac{R_{\rm{max}}-R_{\rm{min}}}{2 R_{\rm{avg}}}$ is the modulation amplitude, derived by calculating the rates for the 
average Earth velocity and varying it by $\pm 15.0$~km/s.  
The $f_{\rm mod}$ spectrum is distinctive, which should provide a helpful discriminant between signal and background.  
The significance of a signal $S$ over a flat background $B$ is then given by 
$sig=\frac{f_{\rm{mod}}\,S}{\sqrt{S+B}}$.  

\begin{figure}[t]
\includegraphics[width=0.48\textwidth]{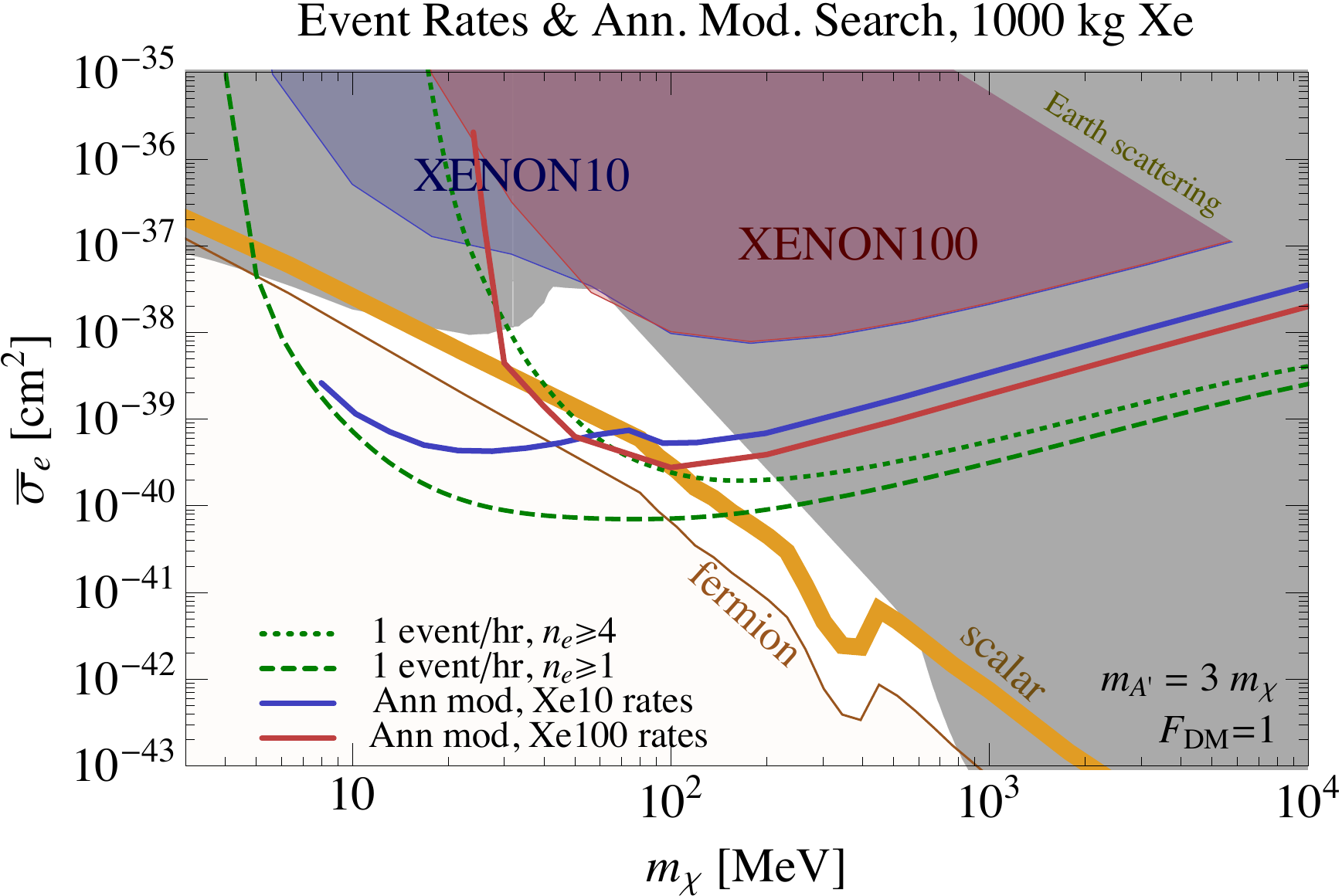}\\
\includegraphics[width=0.48\textwidth]{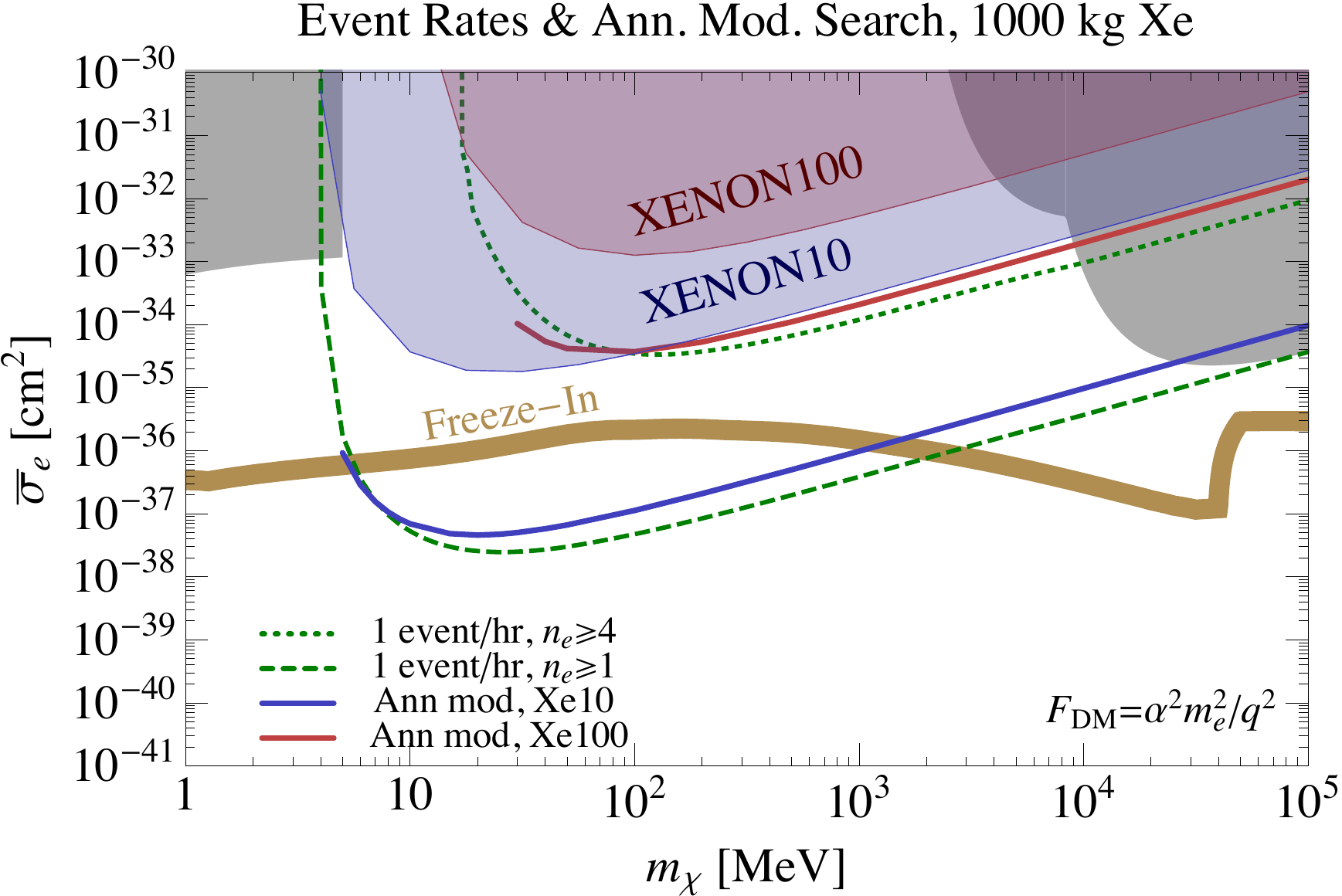}\\
\caption{
Sensitivity reach from an annual modulation analysis with a hypothetical 1000~kg detector and 1-year exposure, assuming the observed 
spectrum and data rate are the same as in XENON10~\cite{Angle:2011th} (solid blue) or XENON100~\cite{Aprile:2016wwo} (solid red).  
DM-electron scattering event rates assuming a 1-electron (4-electron) threshold are shown in dashed (dotted) green.   
Blue (red) shaded regions show our XENON10 (XENON100) limits.  
These lines/regions are overlaid on several simple and predictive benchmark models for DM ($\chi$) scattering off electrons 
via a dark photon $A'$.  {\bf Top:} ($F_{\rm DM}=1$)
A complex scalar obtains the correct relic density from thermal freeze-out (light orange), while a fermion, which obtains its  
correct relic abundance from an initial asymmetry, must have $\overline\sigma_e$ above the dark brown line 
(assuming no additional annihilation channels) 
to avoid indirect-detection constraints~\cite{Madhavacheril:2013cna,Ade:2015xua,Essig:2013goa}.  
{\bf Bottom:} ($F_{\rm DM} = \alpha^2m_e^2/q^2$) 
Fermion DM coupled to an ultralight mediator $A'$ obtains the correct relic density from freeze-in (thick brown line).  
Gray regions show constraints as in~\cite{Essig:2015cda}, updated on the top plot with data from MiniBooNE~\cite{Aguilar-Arevalo:2017mqx} and  BaBar~\cite{Lees:2017lec}.  
Due to earth-scattering effects~\cite{Emken:2017erx}, no XENON10/100 limit exists in the top right region.  
\vspace{-5mm} 
\label{fig:eventrates}
}
\end{figure}

To demonstrate the power of an annual modulation search, we imagine that a future detector with 1000~kg-years of exposure
observes the {\it same} S2-only event rate and spectrum as observed in XENON10 data, $R_{\rm Xe10}$.  
Requiring the signal rate to be less than the observed event rate yields the same constraints as with XENON10 data, 
$\overline{\sigma}_{e, {\rm Xe10}}$.  
However, an annual modulation analysis would potentially see a signal of high statistical significance, and in the absence of one 
a fraction of the observed event rate must be background.  
Requiring the significance of the annual modulation signal to be less than $sig$, the expected sensitivity is 
\beq
\overline{\sigma}^{\rm mod}_e = \frac{sig \times \overline{\sigma}_{e,{\rm Xe10}}}{f_{\rm mod} \sqrt{R_{\rm Xe10} \times exposure}}\,.
\eeq
We calculate $\overline{\sigma}^{\rm mod}_e$ for bins of $n_e=0.5-1.5,~1.5-2.5,\ldots$ and show with a blue line the best sensitivity 
across all bins in Fig.~\ref{fig:eventrates} for $sig = 1.65$ (90\%~CL) (see Appendices for sensitivities from each bin). 
Similarly, a red solid line shows $\overline{\sigma}^{\rm mod}_e$ assuming the future observed rates/spectrum 
correspond to the current XENON100 rate/spectrum. 
We overlay these lines on the DM benchmark models discussed above. 
While hypothetical, this analysis emphasizes the power of an annual modulation analysis.

\mysection{Large Event Rates} 
To further emphasize the importance of understanding the electron recoil events in xenon TPCs, we show 
the expected event rates in Fig.~\ref{fig:eventrates} for a 1000~kg detector for two thresholds, $n_e\ge 1$ and $n_e\ge 4$.  
We see that the benchmark DM models predict large event rates.  For example, Dirac fermion DM coupled to the $A'$ 
that obtains its abundance from an initial asymmetry could produce about one event every two seconds at LZ.  
This underscores the point that while there are several sources of backgrounds that can produce single- or few-electron events, 
a large event rate can be consistent with a DM signal and should not be simply written off as a detector curiosity.  

\mysection{Conclusions} 
We derived new constraints on DM-electron scattering, improving upon the previous bound, and 
showed spectra for the expected number of electrons and the modulation amplitude.  
While there are several possible detector-specific origins of the observed XENON10/100 events, in principle almost 
all the observed 
events could originate from DM-electron scattering without coming into conflict with other existing DM constraints.
This is not the case when interpreting these events  
as arising from few-GeV DM recoiling elastically off nuclei~\cite{Angle:2011th,Aprile:2016wwo}, which is excluded by existing 
results from {\it e.g.}~LUX~\cite{Akerib:2016lao} and CDMSlite~\cite{Agnese:2015nto}.  
Moreover, simple and predictive DM benchmark models predict large event rates in current and future xenon TPCs.  
An expanded and dedicated effort by the xenon collaborations to understand the origin of their low-energy electron recoil data 
is thus imperative and well worth the effort.

\vspace{-4mm}
\section{Acknowledgments} \vskip -3mm
We would like to thank especially Aaron Manalaysay and Peter Sorensen for many insightful discussions.  
We also thank Ran Budnik, Daniel McKinsey, Matt Pyle, and Jingke Xu for useful discussions. 
We are also very grateful to Jeremy Mardon for contributions at the beginning of this project as well as many useful discussions.
R.E.~is supported by the DoE Early Career research program DESC0008061 and through a Sloan Foundation Research Fellowship. 
T.-T.Y.~is also supported by grant DESC0008061. 
T.V. is supported by the European Research Council (ERC) under the EU Horizon 2020 Programme (ERC-CoG-2015 - Proposal n. 682676 LDMThExp), by the PAZI foundation, by the German-Israeli Foundation (grant No. I-1283- 303.7/2014) and by the I-CORE Program of the Planning Budgeting Committee and the Israel Science Foundation (grant No. 1937/12). T.-T.Y. thanks the hospitality of the Aspen Center for Physics, which is supported by National Science Foundation grant PHY-1066293, where part of this work was completed.

\begin{appendix}

\section{APPENDIX}

Here we provide additional details to the calculations described in the main text. We also show spectra plots for additional DM 
masses, as well as the XENON10/XENON100 limits and the prospects for an annual modulation analysis from each PE bin.  
For completeness, we also show the expected daily modulation of the signal rate due to the Earth's rotation. 
\vspace{5mm}

\mysection{Theoretical Rates}
We first quote additional formulas that are required for the rate calculation (see also~\cite{Essig:2012yx,Essig:2015cda}).  
The velocity-averaged differential ionization cross section for electrons in the $(n,l)$ shell is given in Eq.~(\ref{eq:ionization-sigma}).  
The full expression for $v_{min}$ is 
\beq
v_{min}=\frac{\left(|E_{\rm{binding}}^{nl}|+E_{er}\right)}{q} +\frac{q}{2m_\chi}\,,
\eeq
where $E_{\rm{binding}}^{nl}$ is the binding energy of the shell and $q$ is the momentum transfer from the DM to the electron. 
The form factor for ionization of an electron in the $(n,l)$ shell with final momentum $k'=\sqrt{2m_e E_{er}}$ is given by 
\bea
|f_{ion}^{nl}(k',q)|^2=\frac{4k^{\prime3}}{(2\pi)^3}\sum_{l' L}(2l+1)(2l'+1)(2L+1)&&\nonumber\\
\times\left[\begin{matrix} l&l'&L\\0&0&0\end{matrix}\right]^2
 \left|\int r^2dr R_{k'l'}(r)R_{nl}(r)j_L(qr) \right|^2,&&
\eea
where $[\cdots]$ is the Wigner 3-$j$ symbol and $j_L$ are the spherical Bessel functions. We solve for the radial wavefunctions $R_{k'l'}(r)$ of the outgoing unbound electrons taking the radial Schr{\"o}dinger equation with a central potential $Z_{\rm{eff}}(r)/r$. This central potential is determined from the initial electron wavefunction by assuming that it is a bound state of the same potential. We include the shells listed in Table~\ref{table:shells}.

\begin{table}[b]
\begin{center}
\begin{tabular}{|c||c|c|c|c|c|c|}
\hline
 Shell & $5p^6$&$5s^2$ &$4d^{10}$& $4p^6$& $4s^2$\\ \hline
 Binding Energy [eV] &12.4 & 25.7 &  75.6 & 163.5 &  213.8\\ \hline
 Photon Energy [eV] & -- & 13.3 & 63.2  & 87.9 & 201.4\\ \hline
 Additional Quanta & 0 & 0 & 4 & 6-10 & 3-15\\ \hline
\end{tabular}
\caption{Xenon shells and energies. ``Photon energy'' refers to energy of de-excitation photons for outer-shell electrons 
de-exciting to lower shells. 
This photon can subsequently photoionize, creating additional quanta.  
The range of additional quanta takes into account that the higher energy shell may have more than one available lower energy shell to de-excite into. For our limits, we take the minimum of this range.
\vspace{-6mm}
}
\label{table:shells}
\end{center}
\end{table}%

\vspace{5mm}
\mysection{Electron and Photoelectron Yields}
\begin{figure*}[htbp]
\includegraphics[width=0.48\textwidth]{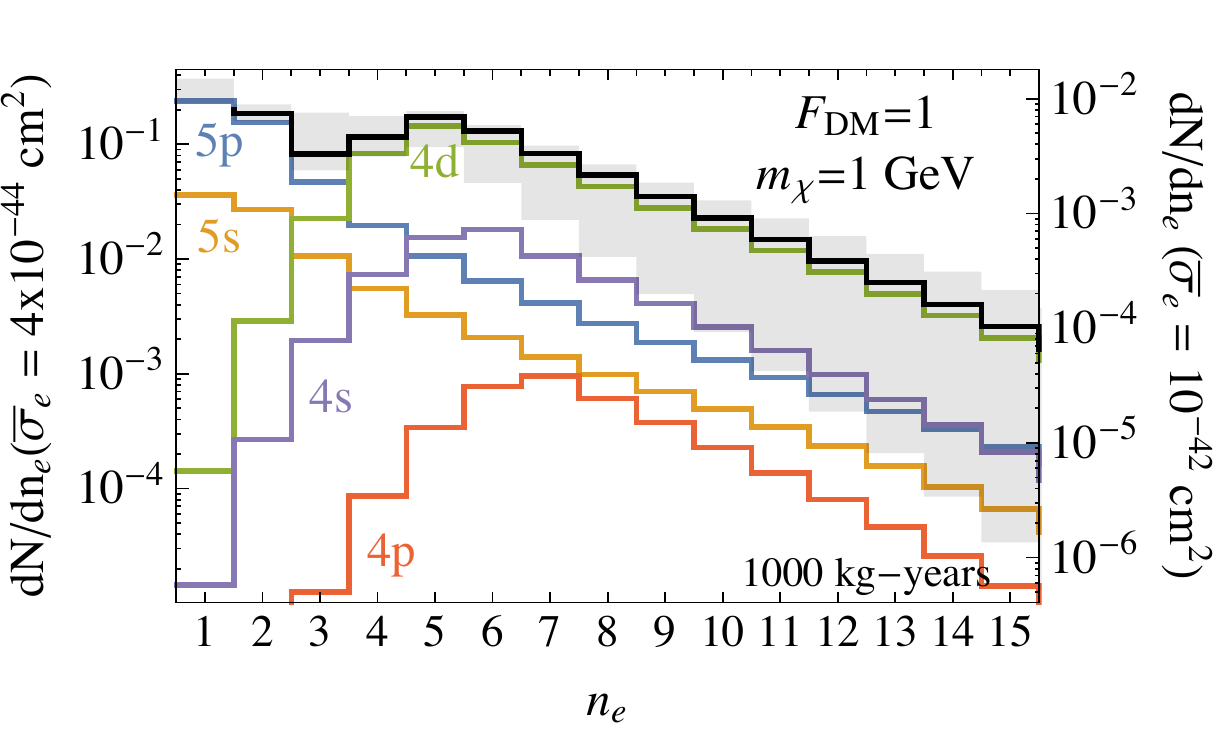}
~~ \includegraphics[width=0.48\textwidth]{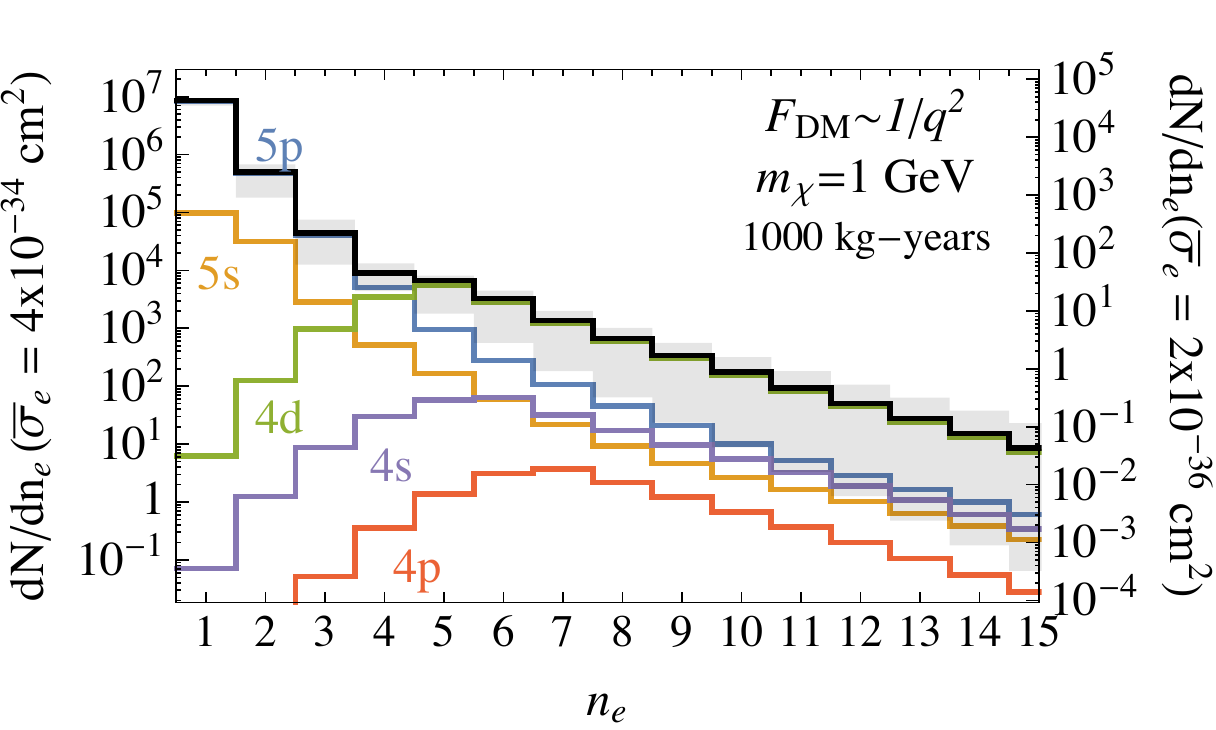}
\vskip -2mm
\includegraphics[width=0.48\textwidth]{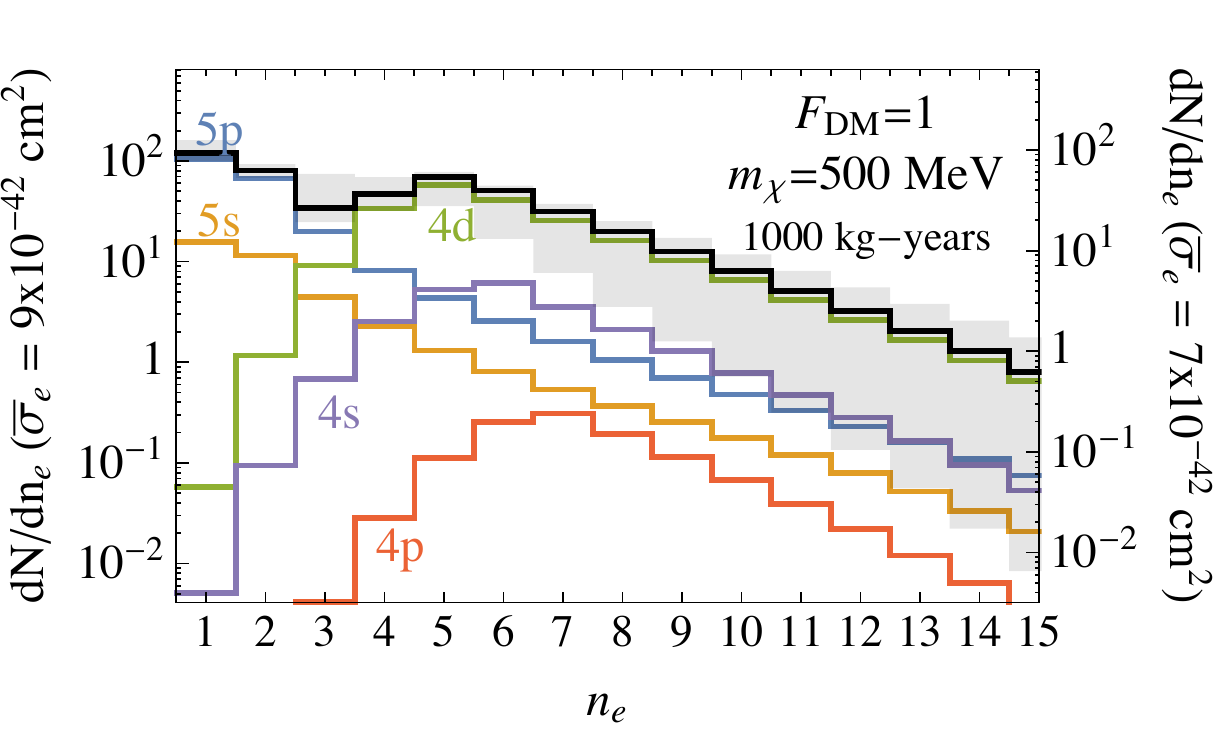}
\includegraphics[width=0.48\textwidth]{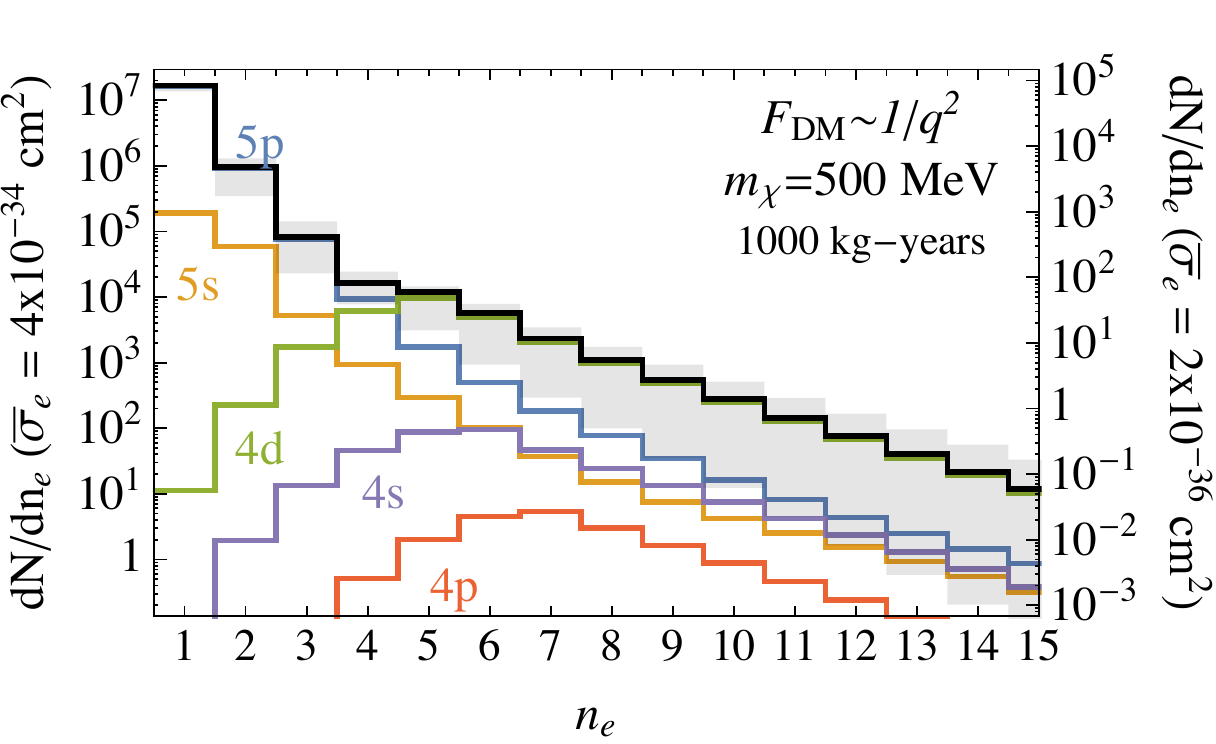}
\caption{Expected number of events as a function of number of electrons observed for 1000 kg-years of xenon. The left-axis sets $\overline\sigma_e$ to the maximum allowed value by current constraints while the right-axis sets $\overline\sigma_e$ to the predicted value for a freeze-out (freeze-in) model for $F_{\rm DM}=1(\alpha^2m_e^2/q^2)$, respectively. The different colored lines show the contributions from the various xenon shells while the gray band encodes the uncertainties associated with the secondary ionization processes.}
\label{fig:spectra}
\end{figure*}
We provide additional details to convert the recoiling electron's recoil energy into a specific number of electrons. 
The relevant quantities are
\bea
E_{\rm er}&=&(n_\gamma+n_e)W\nonumber \,, \\
n_\gamma&=&N_{\rm ex} + f_R N_i\,, \\
n_e&=&(1-f_R)N_i\,.\nonumber
\eea
$E_{\rm er}$ is the amount of deposited energy from the primary electron, which results in a number of observable electrons, $n_e$, unobservable scintillation photons, $n_\gamma$, and heat. 
$W$ is the energy needed to produce a single quanta (photon or electron). 
We take $W=13.8\pm0.9$ eV as the average energy~\cite{W13point8}. 
$E_{\rm er}$ can create both a number of ions, $N_{i}$, and a number of excited atoms $N_{\rm ex}$, where $N_{\rm ex}/N_i\simeq 0.2$ at energies above a keV \cite{W13point8,PhysRevB.76.014115}. 
We take into account the possibility that the primary electron and secondary ions can recombine, which is described by a modified Thomas-Imel recombination model ~\cite{PhysRevA.36.614}, and assume that the fraction of ions that can recombine, $f_R$, is effectively zero at low energy.
This implies that $n_e=N_i$ and $n_\gamma=N_{\rm ex}$. 
The fraction of initial quanta observed as electrons is given by $f_e=(1-f_R)/(1+N_{\rm ex}/N_i)\simeq 0.83$~\cite{Sorensen:2011bd}. 
To capture the uncertainty in $f_R, W$, and $N_{\rm ex}/N_i$, we calculate the rates and limits varying these parameters over the ranges $0<f_R<0.2,~0.1<N_{\rm ex}/N_i<0.3$, and $12.4 < W<16$ eV.  For our fiducial values, we set $f_e=0.83,~f_R=0,~W=13.8$ eV.

For each primary electron with energy $E_{\rm er}$, we assume that there are  additional $n^{(1)}$=Floor($E_{\rm er}/W$) quanta created. 
Furthermore, we assume that the photons associated with the de-excitation of the next-to-outer shells ($5s, 4d, 4p, 4s$),  which have energies (13.3, 63.2, 87.9, 201.4) eV, can photoionize to create an additional $n^{(2)}$=$(n_{5s},n_{4d},n_{4p},n_{4s})$=(0, 4, 6-10, 3-15) quanta, respectively (see Table~\ref{table:shells}). The range in values for the $4p$ and $4s$ shells takes into consideration that there may be more than one outer-shell electron available that can de-excite down to them.  For example, if the $4d$ shell de-excites to $4p$, 6 additional quanta are created, while if the $5s$ shell de-excites to $4p$, it would create 10 additional quanta.  For our fiducial values, we take the lower number of quanta to be conservative. However, the choice of the number of additional quanta only affects $n_e>6$, and even here the difference in event rate is smaller than the uncertainties due to the modeling of the secondary ionization.

The total number of electrons is given by $n_e=n_e'+n_e''$, where $n_e'$ is the primary electron and $n_e''$ are the secondary electrons produced.
$n_e'=0$ or $1$ with probability $f_R$ or $1-f_R$, respectively, while $n_e''$ follows a binomial distribution with $n^{(1)}+n^{(2)}$ trials and success probability $f_e$. 

Given this conversion from $E_{er}$ into $n_e$, we can calculate the differential rate as a function of number of electrons. In addition to the $m_\chi=100$ MeV spectra shown in the main text, we show the spectra for $m_\chi=500$ MeV and 1 GeV in Fig.~\ref{fig:spectra}. 

\begin{figure}[t]
\begin{center}
\includegraphics[width=0.48\textwidth]{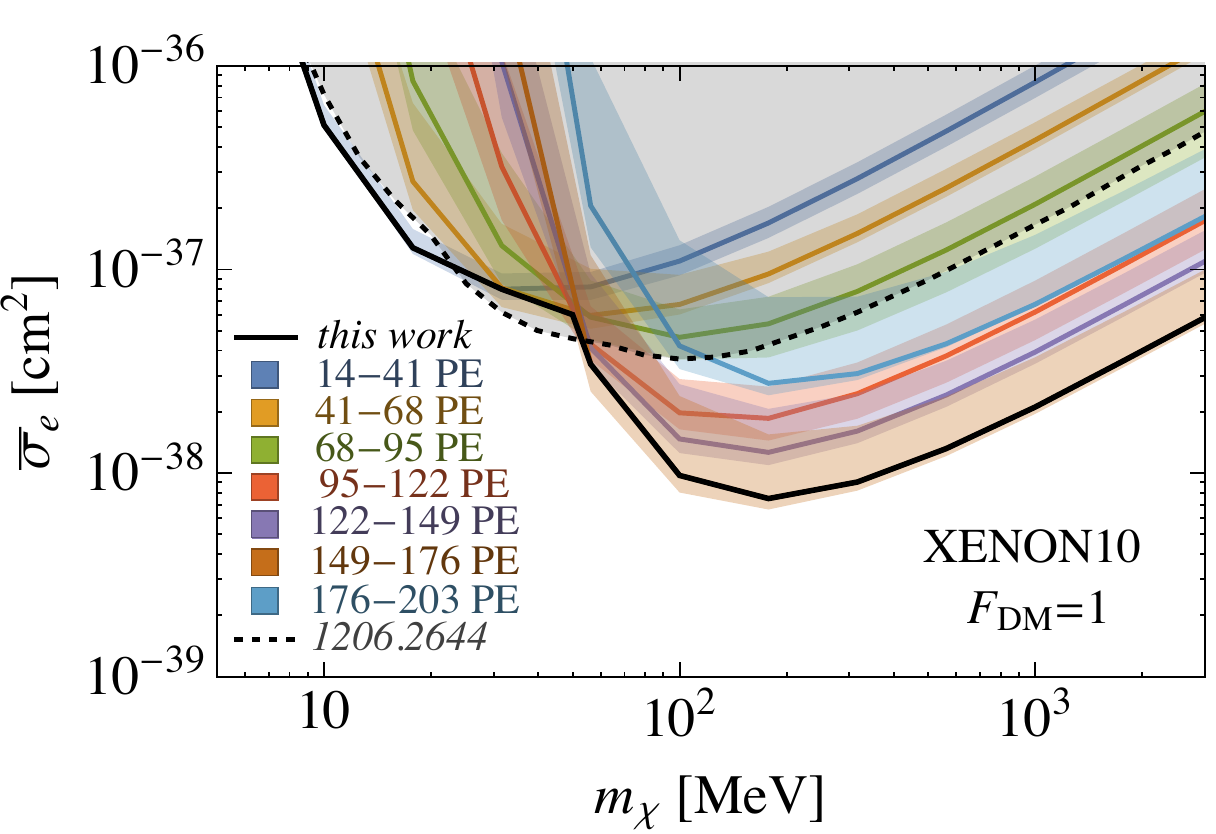} \\
\includegraphics[width=0.48\textwidth]{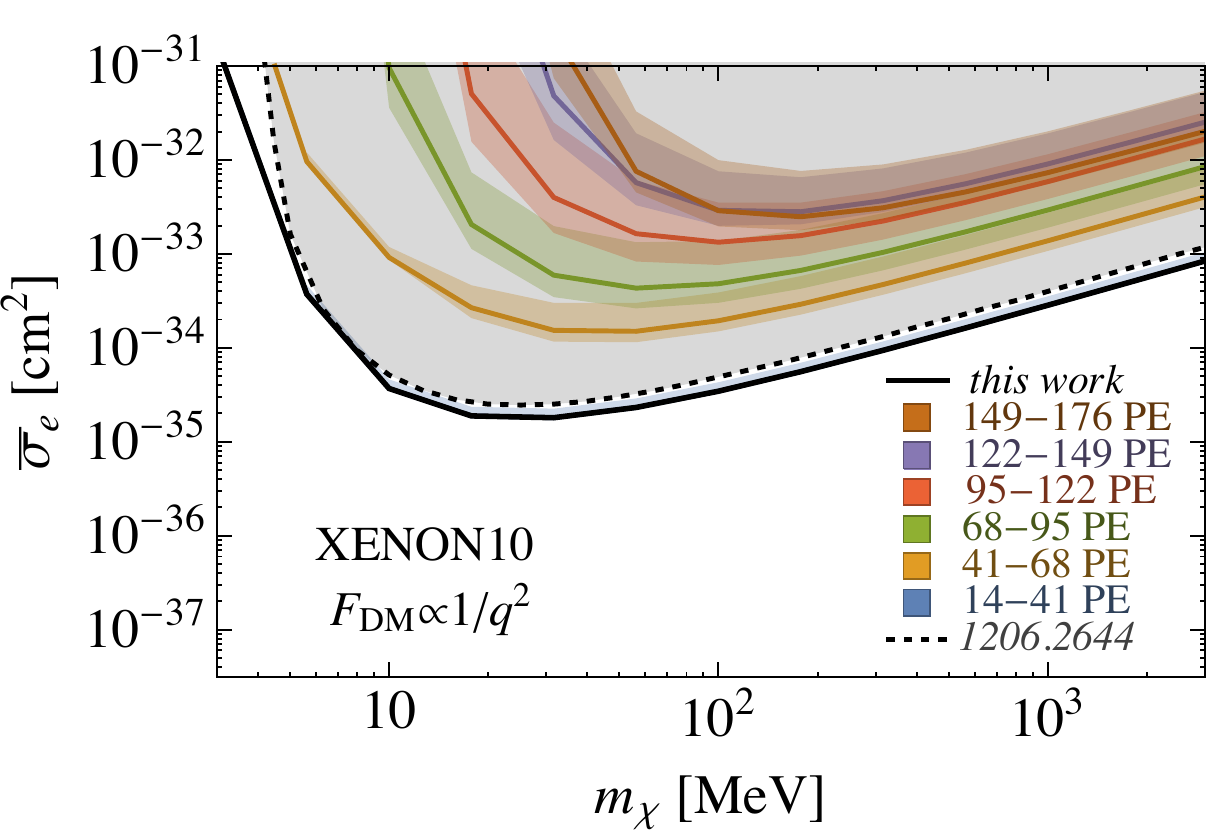}
\caption{New XENON10 limit (black) obtained as described in the text. The colored bands are from the uncertainty in the secondary ionization model. The shaded gray region shows the parameter space previously excluded by the 1, 2, and 3 electron XENON10 data. By including the contributions to the S2 signal from 14PE to 203PE, we see that the limits improve considerably for DM masses above $\sim 50$ MeV for 
$F_{\rm DM}=1$, while there is no improvement due to the momentum suppression for $F_{\rm DM}= \alpha^2m_e^2/q^2$.
}
\label{fig:Xenon10compare}
\end{center}
\end{figure}

\begin{figure}[t]
\includegraphics[width=0.48\textwidth]{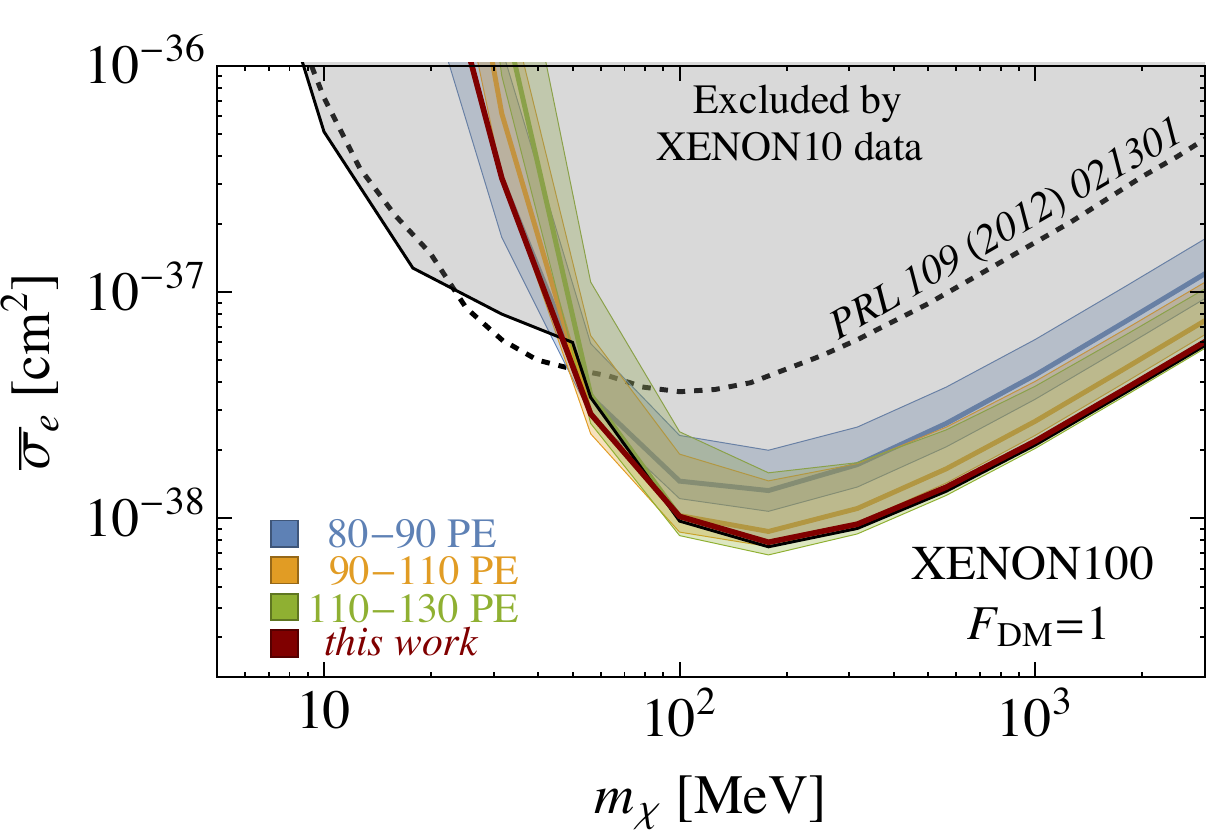} \\
\includegraphics[width=0.48\textwidth]{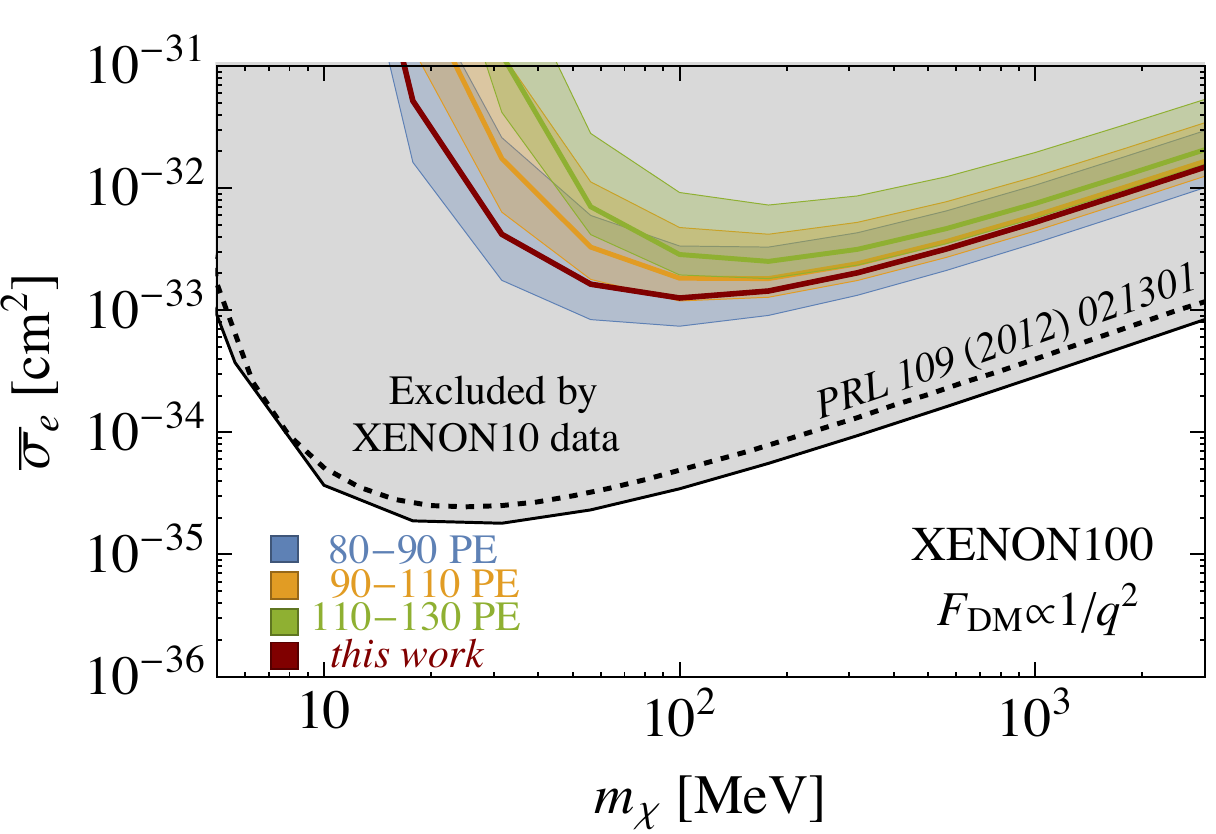}
\caption{New limit obtained using the XENON100 data (red). The XENON100 data starts at 80PE electrons, so we show the individual limits for the 80-90, 90-110, and 110-130 PE bins. The colored bands are from the uncertainty in the secondary ionization model. The shaded gray region shows the parameter space excluded by our updated XENON10 analysis, while the dotted black line shows the 
XENON10 bound from~\cite{Essig:2012yx}. }
\label{fig:Xenon100compare}
\end{figure}

\vspace{5mm}
\mysection{XENON10 and XENON100 constraints for individual photoelectron bins}
In the main text, we show the cross-section limits from the XENON10 and XENON100 data using the fiducial values above. In Figs.~\ref{fig:Xenon10compare},~\ref{fig:Xenon100compare}, we show the individual limits for each PE bin as well as the uncertainty bands due to the secondary ionization model. 

\vspace{5mm}
\mysection{Modulation}
\begin{figure*}[t]
\includegraphics[width=0.47\textwidth]{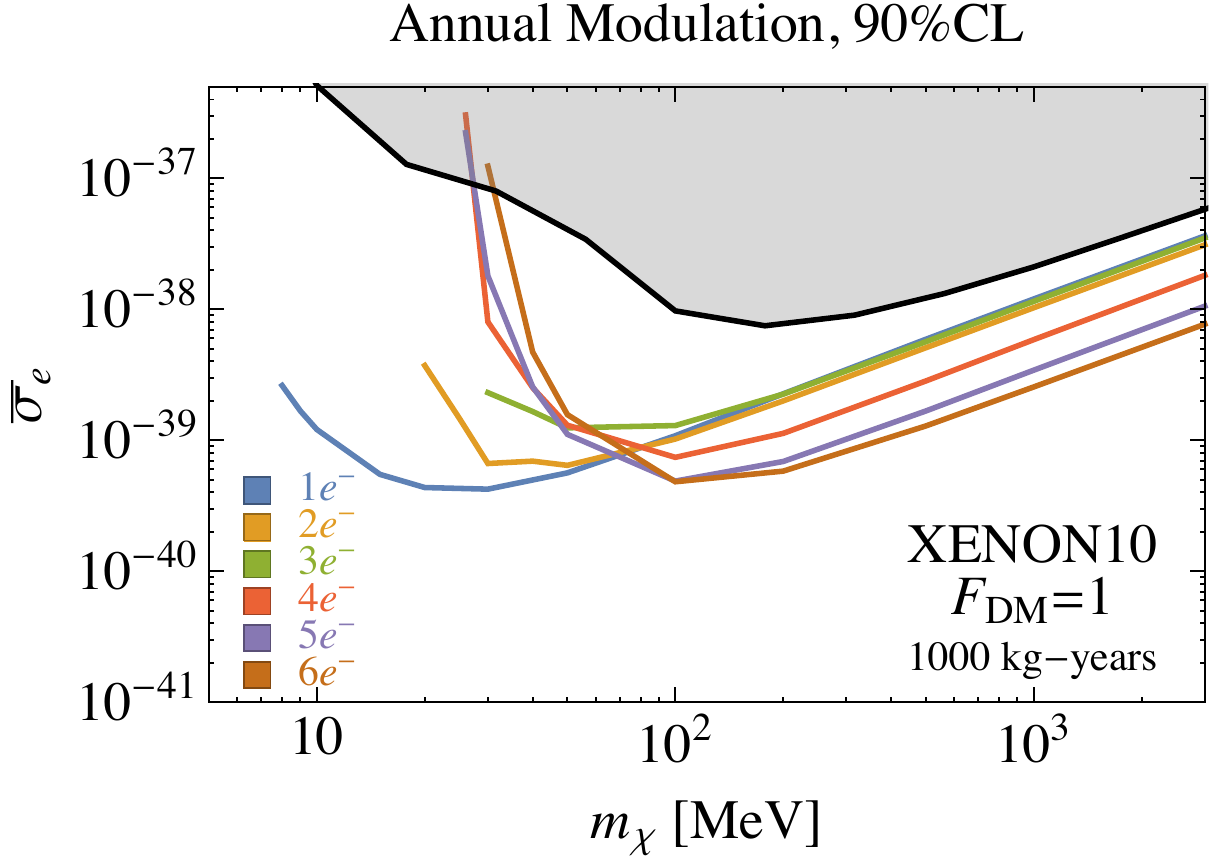}
\includegraphics[width=0.47\textwidth]{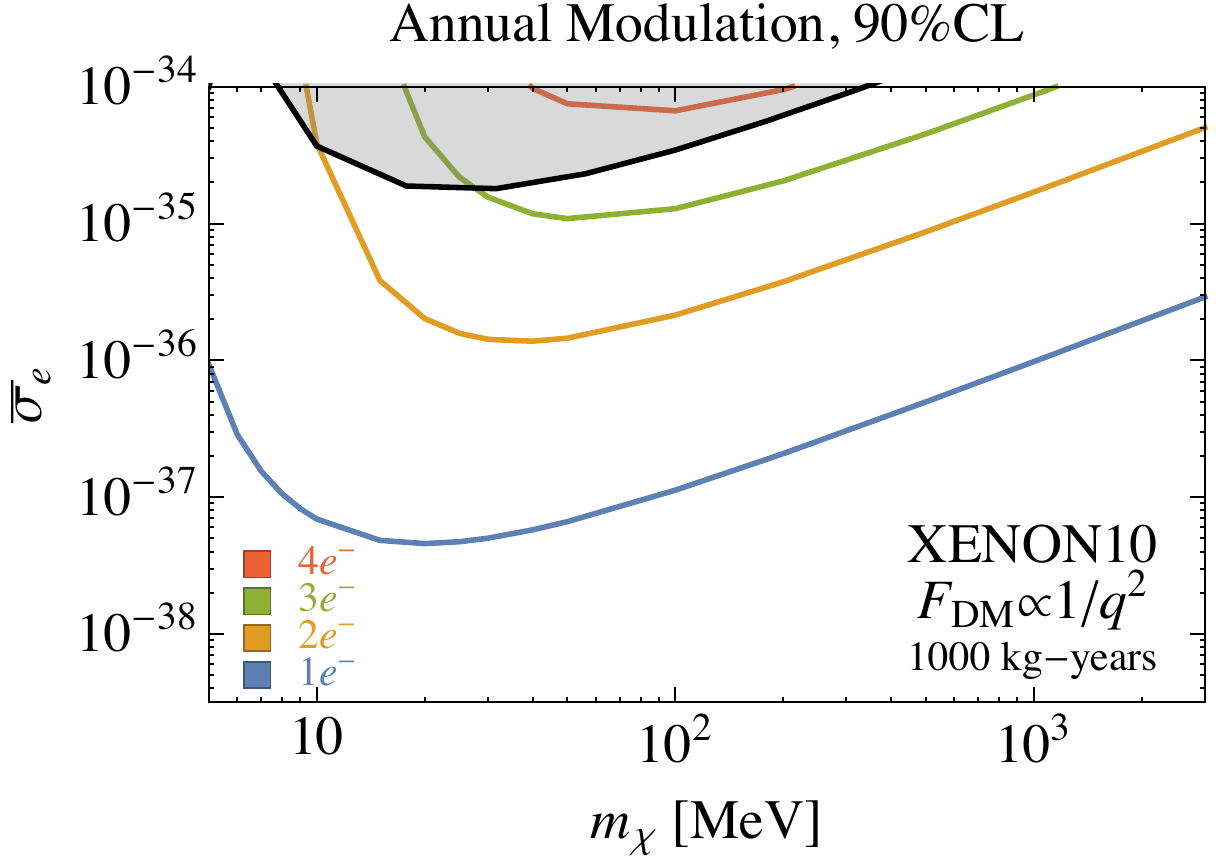}
\includegraphics[width=0.47\textwidth]{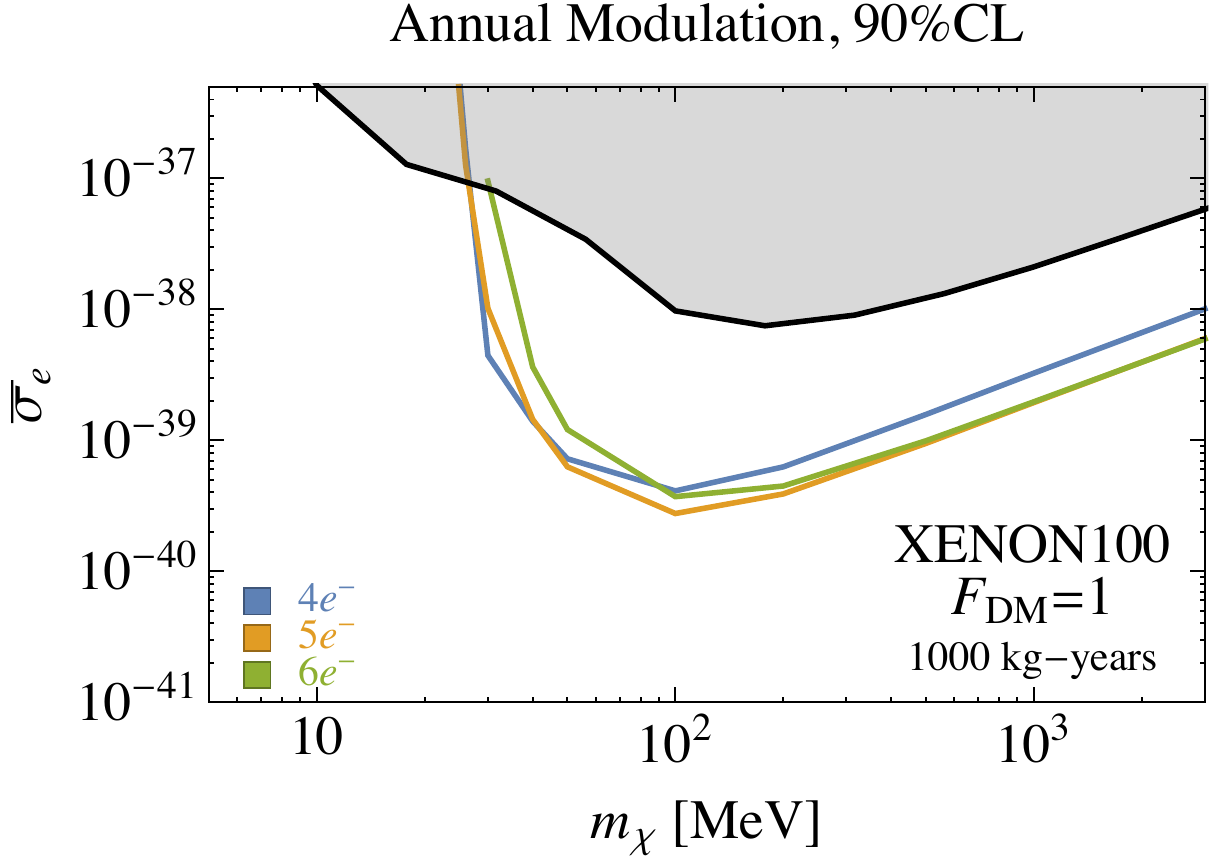}
\includegraphics[width=0.47\textwidth]{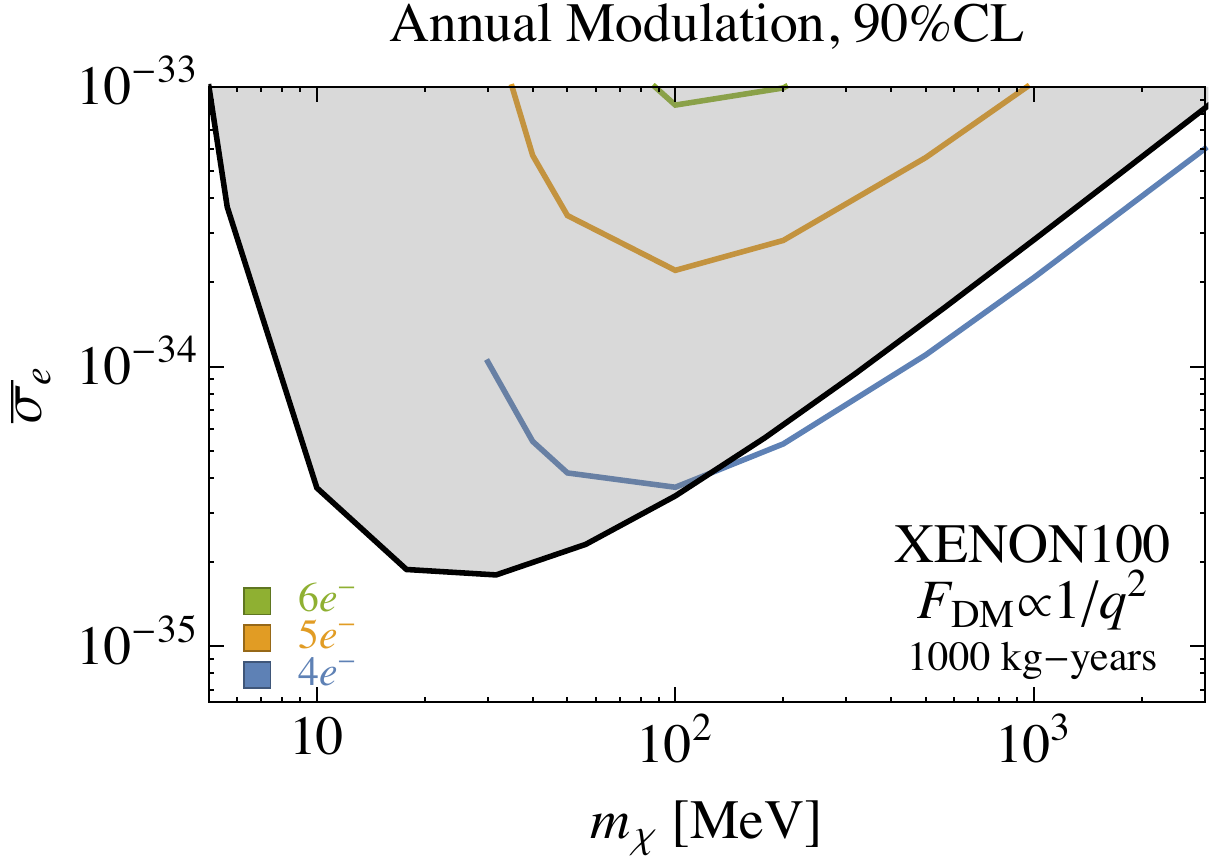}
\caption{Individual bin sensitivities to the 90\% C.L. annual modulation reach for a 1000 kg-year xenon detector. The background rates and spectra are taken to be the XENON10 (XENON100) rates scaled up to 1000 kg-years for the top (bottom) panels (see also text and Fig.~\ref{fig:eventrates}).} 
\label{fig:future-mod}
\end{figure*}
\begin{figure}[h!]
\begin{center}
\includegraphics[width=0.46\textwidth]{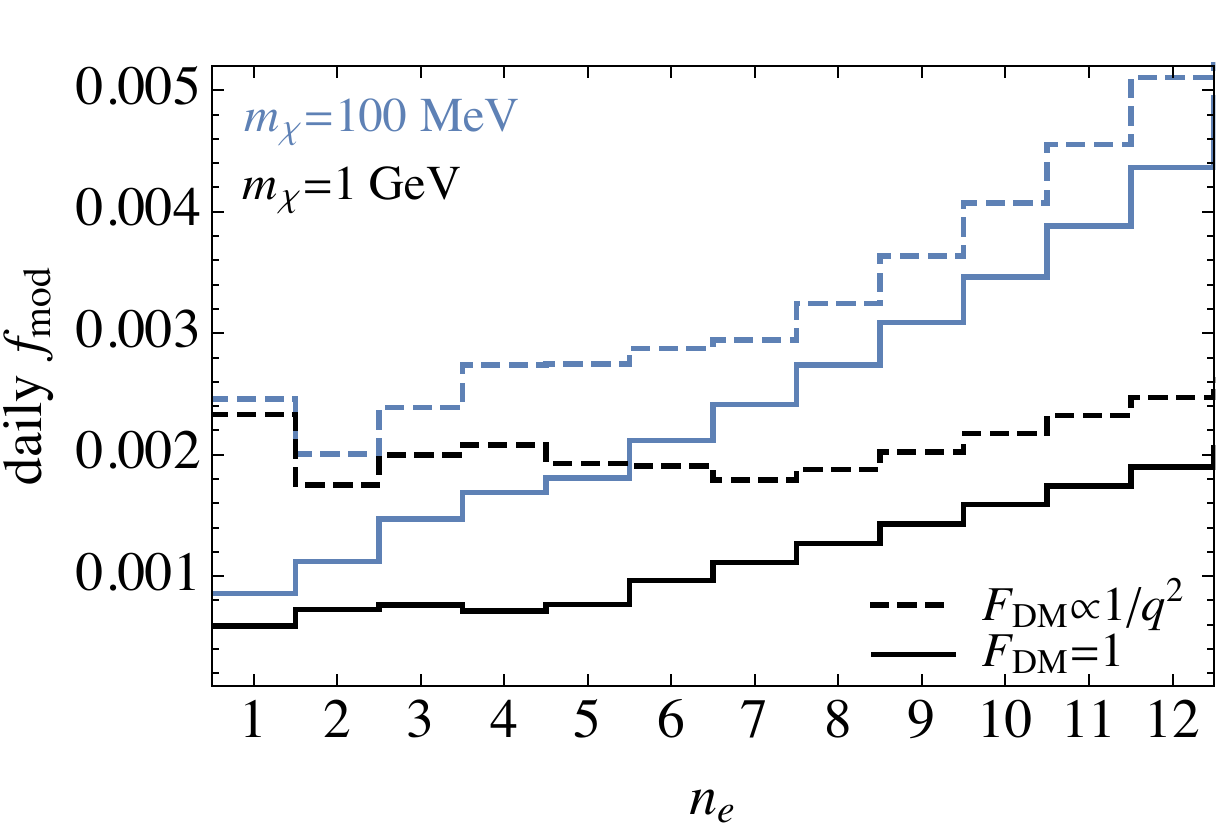}
\caption{Daily modulation amplitude for $F_{\rm DM}=1$ (solid) and $F_{\rm DM} =\alpha^2 m_e^2 /q^2$ (dashed) 
for $m_\chi=100$ MeV (blue) \& 1 GeV (black).}
\label{fig:fmod_vs_Q-daily}
\end{center}
\end{figure}
In Fig.~\ref{fig:eventrates}, we showed how an annual modulation analysis of a hypothetical xenon detector with an exposure of 
1000~kg-years could significantly improve on current constraints even if the background rates are significant.  
In Fig.~\ref{fig:eventrates}, we only showed the best constraints across all individual $n_e$ bins.  
In Fig.~\ref{fig:future-mod}, we show the individual $n_e$ bins.  
Furthermore, for completeness, we also show the daily modulation amplitude due to the Earth's rotation with respect to the DM wind. 
The daily modulation is calculated by modifying the average earth velocity by $\pm 0.23$ km/s to obtain the maximum and minimum rates.
We show the daily modulation fraction in Fig.~\ref{fig:fmod_vs_Q-daily}, where we see that the daily modulation fraction is about an order of magnitude smaller than that of the annual modulation. 

\vskip 4cm
\end{appendix}
\bibliography{LargeRates}

\end{document}